\documentclass[fleqn,usenatbib]{mnras}

\usepackage{newtxtext,newtxmath}

\usepackage[T1]{fontenc}
\usepackage{ae,aecompl}

\usepackage{graphicx}	
\usepackage{amsmath}	
\usepackage[authoryear]{natbib}
\DeclareRobustCommand{\DEL}[3]{#2} 
\DeclareRobustCommand{\DE}[3]{#2}
\DeclareRobustCommand{\DE}[3]{#2}

\title[Fornax bursty SFH]{The bursty star formation history of the Fornax dwarf spheroidal galaxy revealed with the HST}

\author[Rusakov et al.]{
V. Rusakov,$^{1,2,3}$\thanks{E-mail: \href{mailto:rusakov124@gmail.com}{rusakov124@gmail.com}}
M. Monelli,$^{4,5}$
C. Gallart,$^{4,5}$
T. K. Fritz,$^{4,5}$
T. Ruiz-Lara,$^{4,5,6}$
E. J. Bernard,$^{7}$
\newauthor
S. Cassisi.$^{8,9}$
\\
$^{1}$Department of Physics, University of Surrey, Guildford GU2 7XH, UK\\
$^{2}$Cosmic Dawn Center (DAWN)\\
$^{3}$Niels Bohr Institute, University of Copenhagen, Lyngbyvej 2, DK-2100 Copenhagen {\O}, Denmark\\
$^{4}$Instituto de Astrof\'isica de Canarias, E-38200 La Laguna, Tenerife, Spain\\
$^{5}$Departamento de Astrof\'isica, Universidad de La Laguna, E-38205 La Laguna, Tenerife, Spain\\
$^{6}$Kapteyn Astronomical Institute, University of Groningen, Landleven 12, 9747 AD Groningen, The Netherlands\\
$^{7}$Universite C\^ote d\'Azur, Observatoire de la C\^ote d\'Azur, CNRS, Laboratoire Lagrange, France\\
$^{8}$INAF - Astronomical Observatory of Abruzzo, Via M. Maggini, I-64100 Teramo, Italy\\
$^{9}$INFN, Sezione di Pisa, Largo Pontecorvo 3, 56127 Pisa, Italy
}

\date{Accepted 2020 December 24. Received 2020 December 7; in original form 2020 February 22}

\pubyear{2020}

\begin{document}
\label{firstpage}
\pagerange{\pageref{firstpage}--\pageref{lastpage}}
\maketitle


\begin{abstract}
We present a new derivation of the star formation history (SFH) of the dSph galaxy Fornax in two central regions, characterised by unprecedented precision and age resolution. It reveals that star formation has proceeded in sharp bursts separated by periods of low-level or quiescent activity. The SFH was derived through colour-magnitude diagram (CMD) fitting of two extremely deep Hubble Space Telescope CMDs, sampling the centre and one core radius. The attained age resolution allowed us to single out a major star formation episode at early times, a second strong burst $4.6\pm0.4$ Gyr ago and recent intermittent episodes $\sim2-0.2$ Gyr ago. Detailed testing with mock stellar populations was used to estimate the duration of the main bursts and study the occurrence of low-level star formation between them. The SFHs in both regions show common features, with activity at the same epochs and similar age-metallicity relationship. However, clear indications of a spatial gradient were also found, with mean age increasing with radius and star formation episodes being more prolonged in the centre. While some galaxy evolution models predict bursty SFHs in dwarf galaxies and thus a secular origin of the observed SFH cannot be excluded in Fornax, other evidence points to possible mergers or interactions as the cause of its bursty SFH. In particular, we calculated the Fornax orbit relative to the closest dwarfs and the Milky Way and observed a correspondence between the main intermediate-age and young events and peri-passages of Fornax around the Milky Way, possibly indicating tidally-induced star formation.
\end{abstract}




\section{Introduction}

Dwarf galaxies are the most abundant type of galaxies in the Local Group (LG). Their proximity allows for detailed studies with the ability to resolve individual stars. This makes it possible to obtain quantified, resolved star formation histories (SFH) of these galaxies, by means of recreating their observed colour-magnitude diagrams (CMDs) with models, which provides valuable information about the stellar populations present in the galaxies and when they were formed. This information is important to understand how their evolution is affected by both internal and external factors, such as supernova feedback, host dark matter structure properties, UV-reionisation, tidal effects of a host galaxy and interaction with other dwarfs \citep[e.g. LCID project in][and references therein]{Gallart2012}. Different methods and techniques employed to quantify SFHs of stellar populations are described in \cite{Tolstoy1996}, \cite{Gallart1999}, \cite{Hernandez1999}, \cite{Dolphin2002}, \cite{Aparicio2004}, \cite{Aparicio2009}, \cite{Cignoni2010}.

In this work we investigate the SFH of the Fornax dwarf spheroidal (dSph) galaxy. Typical dSph galaxies, also referred to as early-type, are characterised by low gas content in terms of neutral and ionised hydrogen and, consequently, dominated by old stars \citep[e.g. Cetus and Tucana dSphs in][]{Monelli2010a, Monelli2010b}, as opposed to late-type dwarf irregular (dIrr) systems that preserve gas for longer and exhibit younger stellar populations \citep[e.g. IC 1613 in][]{Skillman2014}. This classification is based only on the current properties of dwarfs and complemented by a transition type (dT) with mixed properties of the former two types \citep[e.g. LGS 3 in][]{Hidalgo2011}. Yet \cite{Kormendy1985} suggested that most early-type galaxies are not originally different from the late-type, but that their evolution diverged after their gas was stripped by external processes. This was supported by the evidence of complicated and lasting SFHs of some dSphs (e.g. Carina dSph in \citealp{Hurley1998}; Leo I dSph in \citealp{Gallart1999} and \citealp{RuizLara2020b}), which retained their gas through the intermediate ages. Fornax, too, has demonstrated signs of an extended SFH, with stars formed at old- (> 10 Gyr ago), intermediate- (1-10 Gyr ago) and, possibly, young-age (< 1 Gyr ago) epochs \citep{delPino2013}. Based on their full SFHs \cite{Gallart2015} proposed a classification of dwarf galaxies into ``fast'' and ``slow'' types, where the first form most of their stars in the first few Gyr of evolution, while the latter type experience more extended star formation. Interestingly, these two types do not correspond directly with dSph and dIrr galaxies, a classification that is based in the current morphology. In particular, at least three dSph satellites of the Milky Way (MW), Leo I, Carina and Fornax, are classified as slow dwarfs. The inferred past locations with respect to the MW or M31 of fast and slow dwarfs in the Local Group led \citet{Gallart2015} to hypothesize that the conditions of the environment in which these dwarfs {\it formed} could be an important factor imprinting their subsequent evolution.

Fornax was discovered by \cite{Shapley1938} when studying photographic plates of the constellation of the same name. One of the peculiar findings about the galaxy is that it has 5 globular clusters (GCs, discovered by \citealp{Baade1939}, \citealp{Shapley1939}, \citealp{Hodge1961a}; a discovery of the sixth star cluster was reported in \citealp{Wang2019a}), which makes Fornax one of the few LG galaxies hosting GCs. It is also one of the most luminous LG dwarfs with $M_V=-13.4\pm0.3$ \citep{Irwin1995}, and a stellar mass of $M_{\star}=20\times10^6 \; M_\odot$ (assuming mass-to-light $M/L=1$, \citealp{McConnachie2012}).

The recent estimate of the distance modulus of Fornax $\mu~=~20.818\pm0.015$(statistical)$\pm0.116$(systematic) ($\sim146$ kpc) by \cite{Karczmarek2017} was obtained by using mean near-infrared magnitudes of RR Lyrae (RRL) stars, which had the advantage of being less affected by extinction over studies performed with the same and other distance indicators in visual bands. In any case, estimates from different studies agree well within their errors (see \citealp{Pietrzy2007} and references in \citealp{Karczmarek2017}). The extinction in the direction of Fornax is $E(B-V)=0.021$, as per \cite{Schlegel1998} extinction maps recalibrated by \cite{Schlafly2011}. 

The proximity of Fornax makes it possible to resolve individual stars and carry out detailed wide-field studies of its structure and stellar content. The structure of Fornax was first studied by \cite{Hodge1961b} with photographic plates, and later by \cite{Hodge1974}, \cite{Eskridge1988}, \cite{Demers1994}, who confirmed and provided details on the initial findings. \cite{Hodge1961b} identified, using star counts, that the ellipticity increased at larger radii and determined the direction of the major axis. \cite{Irwin1995} observed Fornax to have an asymmetric stellar distribution, with a larger density on the east of the major axis and a sparseness on the west, with the innermost regions being the least elliptical. The low ellipticity at the centre is likely due to the misalignment of the major axis of the distribution of the young main sequence (MS) stars from the rest of the galaxy, as shown by the CCD wide-field photometry of 1/3 deg$^2$ central field of Fornax in \cite{Stetson1998}. The authors also observed a generally extended distribution of the old-age RRL stars, and a more concentrated intermediate-age component, represented by the red clump (RC) stars, which also coincided with the dense region on the east-side of the major axis. Current values of the ellipticity and position angle are estimated to be $e=0.31\pm0.01$ and $PA=41.5\pm0.2^{\circ}$ (25 deg$^2$ VST/ATLAS photometry of \citealp{Bate2015}). In the same study the galaxy's tidal radius was estimated to be $r_t=69.7\pm0.3$ arcmin (with core radius $r_c=14.6\pm0.1$ arcmin) from the centre of Fornax located at $\alpha_{2000}=2^h39^m51^s$, $\delta_{2000}=-34^{\circ}30'39''$.

No clear evidence of HI or H$\alpha$ gas in Fornax was found so far \citep{McConnachie2012}, although it shows indications of young stellar populations. A small sub-structure at $r\simeq0.3^{\circ}$ to the south-east, along the minor axis, was identified at approximately one core radius by \cite{Coleman2004} and found to contain more young stars (< 2 Gyr old), than a control field in Fornax investigated by \cite{Olszewski2006}. Another overdensity was located by \cite{deBoer2013} at $r\simeq0.3^{\circ}$ to the east of the centre of Fornax. By calculating the SFH of the region, it was argued that the structure contained stars as young as 0.1-0.3 Gyr old. Later, two more overdensities were found by \cite{Bate2015} to the west and south-south-west of the centre, which were reported to be similar to the previous two.

Although the questions regarding the cause of the young stellar component in the sub-structures of Fornax yet remain unanswered, there has been sufficient evidence to claim that recent star formation events (< 1 Gyr ago) took place across the whole central region causing a clear population gradient, with the youngest and most metal-rich stars concentrated in the centre and the oldest and most metal-poor following the most extended distribution \citep{delPino2015, Bate2015, delPino2013, deBoer2012, Coleman2008, Battaglia2006, Saviane2000, Stetson1998}. Among the latter, the most detailed up-to-date study of the structural parameters of individual stellar populations and their radial profiles within the tidal radius was performed in \cite{delPino2015}. The studies by \cite{delPino2013}, \cite{deBoer2012} and \cite{Coleman2008} obtained their results by constructing SFHs of either individual small regions or wide fields in Fornax by using the CMD fitting technique with ground-based photometry. The deepest photometric data, reaching $I\sim24.5$, was used in \cite{delPino2013}, where the authors claimed the resolution of star formation rate (SFR) from $\sim 1.7$ Gyr at the oldest ages to $\sim 0.6$ at an age of 3 Gyr ago.

Additional interest in the Fornax stellar populations was stirred by studies that combined spatial, chemical and dynamical information into their analyses, and discovered complex patterns across the galaxy. For example, \cite{delPino2017} demonstrated these patterns in terms of angular momenta and line-of-sight velocities; \cite{Amorisco2012} and \cite{Hendricks2014} used different approaches to fit stellar populations either based on their metallicity or the whole set of spatial and chemo-dynamic features. These works were built on the spectroscopic measurements analysed and published by \cite{Pont2004}, \cite{Battaglia2006}, \cite{BattagliaStarkenburg2012}, \cite{Walker2009}, \cite{Kirby2010}, \cite{Letarte2010}.

The goal of this study is to investigate two small regions in the centre of Fornax, with little crowding, to obtain their SFH with high accuracy and resolution in age, exceeding that of the previous studies. It is done by using high-precision photometry reaching the oldest main sequence turnoff (MSTO), going substantially deeper than the previous studies, down to F814W$\sim$27 with the space-based HST observations.

In Section~\ref{sec:observations} we present the observations, describe the photometry procedure and artificial stars tests and discuss the derived Fornax CMDs. Then, the SFH procedure is briefly explained in Section~\ref{sec:sfh_procedure}. The SFH results, tests with mock stellar populations and comparison of the results with previous works are shown in Section~\ref{sec:sfh_results}. The main findings in the paper are discussed in Section~\ref{sec:discussion}, which also demonstrates the results of numerical orbit integration for Fornax and speculates about connection of the orbit with the SFH. The work concludes by the summary in Section~\ref{sec:summary}.

\section{Observations} \label{sec:observations}
\subsection{Data overview} \label{sec:data}

\begin{figure}
	\includegraphics[width=\columnwidth]{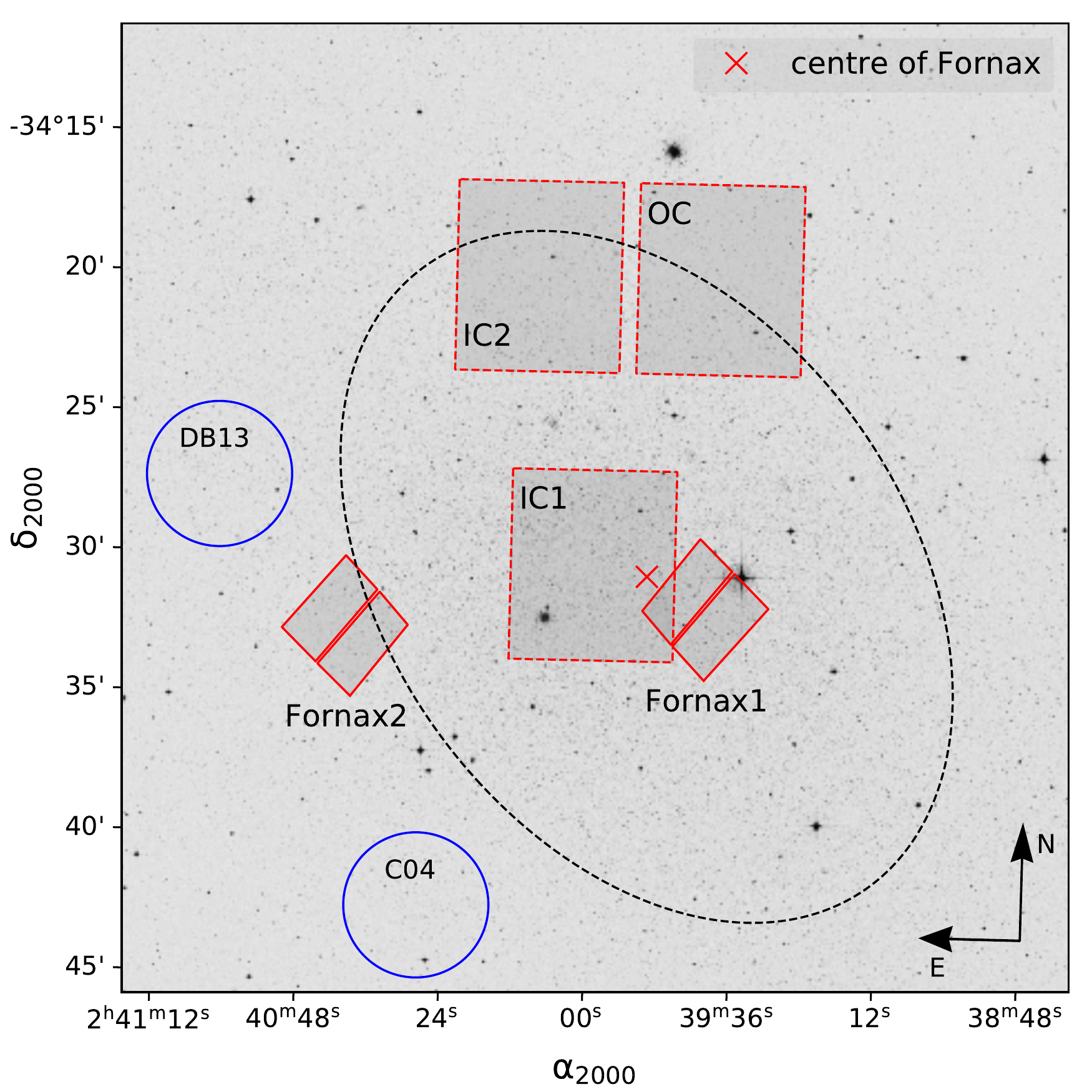}
    \caption{DSS image of the core region of Fornax dSph with camera pointings overlapped. Fornax1 and Fornax2 are rectangular regions in red solid lines. IC1, IC2, OC regions in red dashed lines are from \citet{delPino2013}. Two blue circles of radius 3 arcmin show the locations of two overdensities, discovered by \citet{deBoer2013} and \citet{Coleman2004}, respectively, were adopted from \citet{Bate2015} (two other, to the west and to the south-south-west, discovered by the latter study did not have coordinates specified, but are not expected to overlap with Fornax1 and Fornax2). The dashed ellipse signifies the core radius. The central position of Fornax is marked by the cross. Arrows at the bottom right indicate the directions to the north and east.}
    \label{fig:footprints}
\end{figure}

The observations of the two regions in Fornax at the positions $(\alpha_{2000},\delta_{2000})=(2^h39^m42^s, -34^{\circ}31'43'')$ and $(2^h40^m39^s, -34^{\circ}32'41'')$, hereafter called arbitrarily Fornax1 and Fornax2, were performed with the ACS/WFC camera of the Hubble Space Telescope (HST) as part of the proposal 13435\footnote{Text of the proposal is available at Barbara A. Mikulsi Archive for Space Telescopes at: \url{http://archive.stsci.edu/proposal_search.php?id=13435&mission=hst}.} (PI: M. Monelli) in 2013/14. The two ACS pointings were collected as parallel observations of the WFC3, centred on the GC 4 of Fornax. Given that half of the orbits were taken with a 180$\degr$ rotation, this resulted in two ACS fields separated by $\sim12$ arcmin. The WFC has a field of view of ($202 \times 202$) sq. arcsec and consists of two CCDs stacked side-by-side. The camera took 9 dithered images of each region in F475W and F814W filters, three for each of the 80-, 520- and 850-second exposures. The filters were chosen to cover a large bandwidth, as advised in \cite{Stetson1993}. Table~\ref{tab:datalog} summarises the major image characteristics. The footprints of the two camera pointings are overplotted on top of a DSS image of the core of Fornax\footnote{The background image was taken from the ESO Online Digitized Sky Survey (DSS) at \url{https://archive.eso.org/dss/dss}.} on Figure~\ref{fig:footprints}, together with the footprints from the previous study of SFH in Fornax by \cite{delPino2013}. Fornax1 is a near-central region, located at a radial distance of $r=2.8$ arcmin to the south-west of the galaxy centre, while Fornax2 is close to the core of Fornax, at $r=12.7$ arcmin to the east.

\subsection{Photometric procedure} \label{sec:photometric_procedure}

The photometry was obtained by using DAOPHOT and ALLFRAME software (\citealp{Stetson1987}, \citealp{Stetson1994}). Description of the general procedure and its routines, as well as comparison of DAOPHOT against other software can be found in \cite{Monelli2010a}. Below is the summary of the procedure.

The two chips of the camera were processed separately, as the common field of view was discontinuous due to the gap of $2.5''$ between the chips. The .FLC images (bias-corrected, dark-subtracted, flat-fielded and corrected for the charge transfer efficiency) were used here. They provided correct astrometric positions of the light sources, but their distorted pixels (due to the offset alignment of the ACS instrument within the telescope) resulted in distorted photometry. It was corrected by applying a pixel-area mask.

First, the gaussian centroids of all sources were identified up to 3-sigma level in all images and photometry was performed within 3-pixel apertures, all done with routines in DAOPHOT, which rejected the foreground stars that appeared as sources with saturated peaks. Then, the estimates of the instrumental magnitudes were improved by fitting a Lorentz point spread function (PSF) and subtracting sky values in ALLSTAR, after which coordinate transformations were applied to the images by filter using DAOMASTER. Then the individual images were median-combined to remove cosmic rays and any other spurious sources (such as satellites or bad pixels). 

This clean image was used to search for stellar sources, through two iterations of the process from finding the sources with FIND and evaluating their magnitudes with ALLSTAR. The derived list of stars was adopted as input catalogue for ALLFRAME, which generated an individual catalogue of stars for each image. Then, the process was repeated re-modelling the PSFs using the ALLFRAME catalogues. We adopted the Moffat function with $\beta=1.5$ and $\alpha$ linearly varying across the space in the images, and based on an average sample of 1000 brightest stars (in F814W) for Fornax1 and 620 stars for Fornax2, selected from the stars with refined positions and magnitudes from the previous step. A new stacked median image was generated using updated geometric transformations and the second and final ALLFRAME was run to obtain the final catalogues of stars. Different extended-shape sources (with $\vert$sharpness parameter$\vert \geq 0.1$, see \citealp{Stetson1987}), such as background galaxies, were removed from the catalogues.

Several corrections were performed to calibrate the obtained catalogues. The magnitudes were shifted according to the aperture correction to the standard ACS $0.5''$ aperture. The correction was calculated as the median difference between the estimated magnitudes of the PSF stars and magnitudes of the same stars obtained by using the curvature growth method with DAOGROW \citep{Stetson1990}. The stellar catalogues were calibrated to the VEGAMAG photometric system with the zero-points $ZP_{F475W}=26.150$ and $ZP_{F814W}=25.518$ for each filter, taken from \cite{Bohlin2016}. Additionally, the absolute values of magnitude corrections from \cite{Sirianni2005} $\Delta M_{F475W, inf}=0.087$ and $\Delta M_{F814W, inf}=0.079$ were subtracted from the estimated magnitudes to account for stellar flux at the infinite aperture. The CMDs resulting from the obtained photometry are discussed in Section~\ref{sec:observed_cmd}.

\subsection{Artificial star test} \label{crowding_test}

Photometric errors in the images were estimated by means of the artificial star tests (ASTs), in which artificial stars were used to probe different levels of stellar crowding, as prescribed in \cite{Gallart1996a}. This allowed evaluating the effect of crowding and other various possible errors. This step was essential to simulate the same photometric errors and completeness in the synthetic CMD, as in the observed one (see \ref{sec:synth_cmd}), which was essential for an accurate construction of the SFH.

To perform the tests a synthetic CMD with $5\times10^6$ stars and with stellar evolutionary models and assumptions outlined in Section~\ref{sec:synth_cmd} was created. The synthetic stars were injected into the images at the nodes of a regular grid by using the previously obtained PSF, with the distance between the nodes of ($2R + 0.5$) pixels, such that the artificial stars interfered only with the light profiles of the real stars and not with each another. We used the value $R=10$ pixels (for comparison, average PSF aperture radii were equal to $R_{PSF}=6.6$ for stars observed in Fornax1 and $R_{PSF}=6.2$ in Fornax2). Additionally, we implemented a random change of ($0-0.3\%$) in magnitudes of artificial stars to account for the effect of charge transfer inefficiency of the ACS camera mentioned in \cite{Brown2014}. Given the number of the artificial stars, distance between nodes in the grid and image dimensions, a number of iterations were required to complete the photometric measurement of all artificial stars.

Stars were recovered from the images by employing the same photometric procedure as for the observed photometry (Section~\ref{sec:photometric_procedure}), and the output artificial stars were separated out. The only difference in the photometric procedure this time was the PSF. It was defined in the same way as before, but based on fewer stars, so as to degrade it to a lower quality. It was done to avoid using the same PSF for injecting and recovering the stars, which could underestimate the errors for the artificial stars, compared with the observed. The degraded PSF was defined by using a random fraction of stars of the original PSF star list in the range $80-95\%$.

Most artificial stars were recovered with magnitudes differing from the input magnitudes, with the difference being due to the crowding effects and photometric errors, while some stars were lost, i.e. not recovered or recovered with a magnitude beyond a threshold value. This information was used for simulation of observational effects in the synthetic CMD later.

In conclusion, the fainter stars were recovered with the greatest magnitude difference, i.e. they were prone to larger errors. The photometric errors did not significantly affect the photometry even in the most crowded regions of Fornax1 and Fornax2, as the fraction $\sim99.6\%$ of stars experienced a magnitude change of $\rm \vert \Delta M_{in-out}\vert \leq 0.5$, while for $\sim90\%$ of stars the maximum magnitude difference was only $\rm \vert \Delta M_{in-out} \vert \leq 0.1$ in both magnitudes.

\subsection{Observed CMD} \label{sec:observed_cmd}

\begin{figure} 
	\includegraphics[width=\columnwidth]{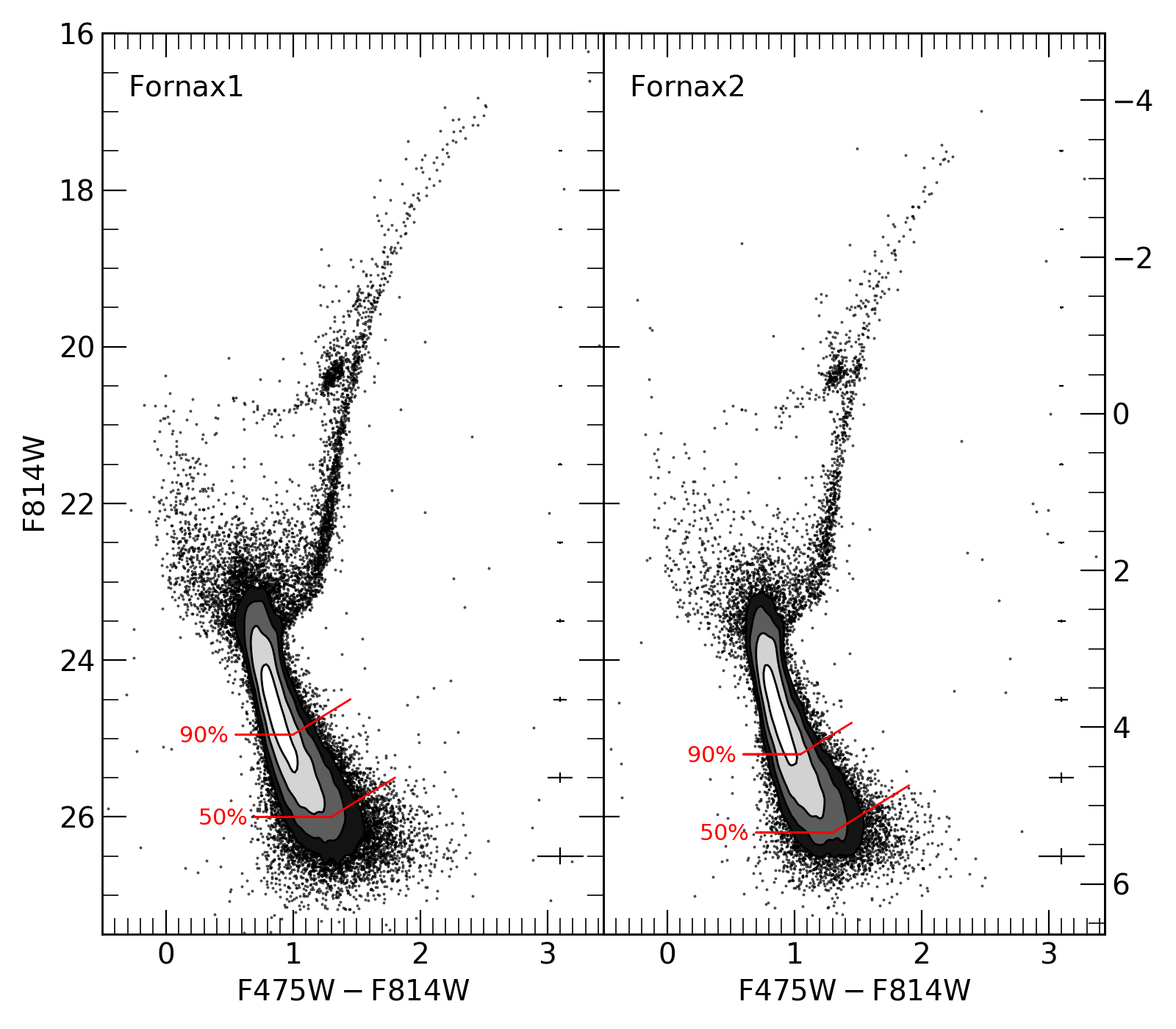}
    \caption{Observed CMDs of Fornax1 (left panel) and Fornax2 (right panel) in the planes of apparent (left axis) and absolute (right axis) magnitudes. Red lines indicate the levels of completeness estimated in the ASTs (top line: 90\% level; bottom line: 50\% level). The photometric errors in colour and magnitude, defined by the typical sigma values of magnitudes at all colours $m_{F475W}-m_{F814W}$ and corresponding magnitudes $m_{F814W}$, are indicated on the right side of the two panels. The grey contours along the MS show the $20, 50, 70, 80\%$ confidence regions of stellar density, respectively.}
    \label{fig:obs_cmds}
\end{figure}

Figure~\ref{fig:obs_cmds} shows the observed CMDs in the planes of the calibrated apparent (left axis) and absolute magnitudes (right axis). The CMDs of Fornax1 and Fornax2 are similar and span $\sim 10$ magnitudes in $m_{F814W}$. They reach $\Delta m_{F814W}\sim1.5$ and $\Delta m_{F814W}\sim1.7$ below the oldest MSTO, to the level of $90\%$ completeness, respectively, which makes this photometry set the deepest among the recent studies of various regions in Fornax (\citealt{deBoer2012, delPino2013}). In addition, the high spatial resolution of the ACS images and the deep photometry in both fields produce very accurate CMDs that reveal various features distinctly. 

There are clear signs of presence of stars of most ages. Young main sequence (MS) stars populate both CMDs at $-0.2 < m_{F475W} - m_{F814W} < 0.4$ and $20.5 < m_{F814W} < 21.5$, though their relative number is larger in Fornax1 than in Fornax2 indicating more active star formation in the recent times towards the central region. The few brightest (youngest) stars are found in Fornax2, possibly suggesting some small spatial scale variation or stochasticity in the latest event of star formation. We clearly observe the oldest MSTO and at least two subgiant branches (SGB; previously seen for example in \citealt{Buonanno1999}) of the intermediate-age and old population that present an interesting complex morphology. For instance, one distinct SGB with intermediate age stars departing from the bulk of stars in the MS at $(m_{F475W} - m_{F814W}) \simeq 0.8$ and $m_{F814W} \simeq 23.2$ and joining the RGB $\simeq$ 0.4 magnitudes brighter and redder. The complex evolution of Fornax is fully represented in the helium-burning phase. The oldest stars show up in the mostly red horizontal branch, which presents few stars bluer than the instability strip, possibly merging with the young MS. However, the most populated feature is the RC, which looks elongated in luminosity, and is well separated from the RGB, strongly suggesting a limited chemical enrichment. Possibly, a few counterparts of young stars are visible, spread above the RC at $1.2<(m_{F475W}-m_{F814W})<1.4$ and $18.5<m_{F814W}<20.0$. The subsequent asymptotic phase (AGB) is also very well defined, with the AGB clump \citep{Gallart1998} showing up at m$_{F814W} \sim$19.3 mag.

Overall, the number density of stars in most CMD regions in the near-central region Fornax1 decreases towards Fornax2, located at a larger distance from the centre, with 41656 and 24058 observed stars, respectively.

\begin{figure}
	\includegraphics[width=\columnwidth]{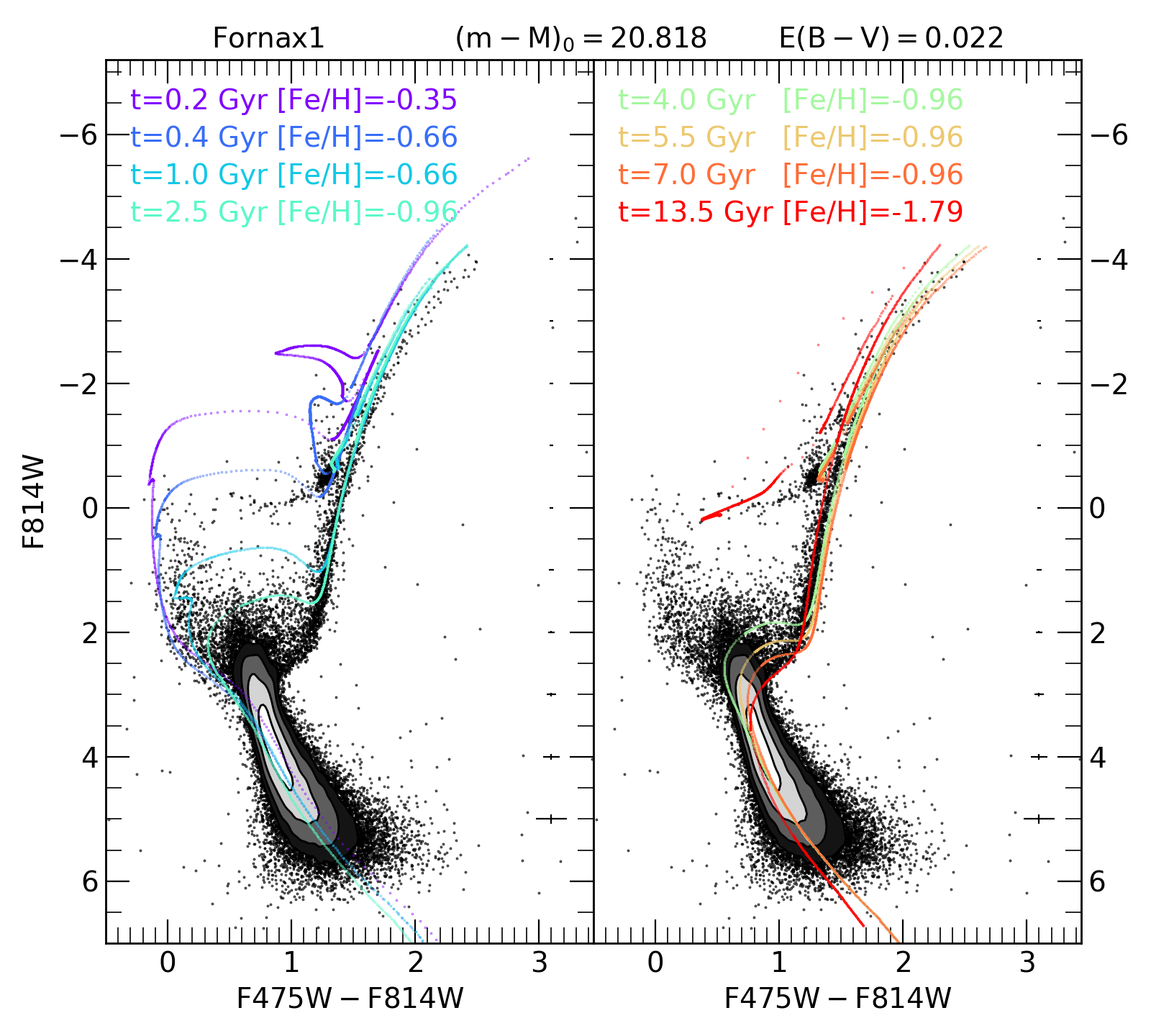}
    \caption{Observed CMD of Fornax1 in the plane of absolute magnitudes with some of the youngest (left panel) and oldest isochrones (right panel) overplotted. The photometric errors in colour and magnitude, defined by the typical sigma values of magnitudes at all colours $M_{F475W}-M_{F814W}$ and corresponding magnitudes $M_{F814W}$, are indicated on the right side of the two panels. The grey contours along the MS show the $20, 50, 70, 80\%$ confidence regions of stellar density, respectively.}
    \label{fig:obs_cmds_iso}
\end{figure}

A more detailed analysis of the stellar populations can be done by fitting model isochrones to the various evolutionary sequences in the CMD for the more populated Fornax1 region. Figure~\ref{fig:obs_cmds_iso} shows the observed CMDs shifted to the plane of absolute magnitudes by using a true distance modulus $\mu=20.818\pm0.015$(statistical)$\pm0.116$(systematic) and extinction $E(B-V)=0.022$ transformed to $A_{F475W}=0.079$, $A_{F814W}=0.037$ for Fornax1, with the transformations from \cite{Bedin2005} ($E(B-V)=0.02$ has been used for Fornax2 in the subsequent analysis). For this comparison we selected some solar-scaled isochrones for appropriate values of age and metallicity from the BaSTI\footnote{The BaSTI library is available at the following web address: \url{http://basti.oa-teramo.inaf.it}} stellar evolution library \citep{Pietrinferni2004}.

The left panel of the figure shows isochrones for intermediate and young ages superimposed on the younger and more metal-rich part of the MS, which is represented by a relatively small number of stars. The blue limit can be approximated by the 0.2 and 0.4 Gyr old isochrones and some other populations can be traced with the 1.0 and 2.5 Gyr old isochrones.

The right panel of Figure~\ref{fig:obs_cmds_iso} shows some intermediate-age and old isochrones superimposed on the CMD of Fornax1. The isochrone at $\rm t=13.5$ Gyr and $\rm [Fe/H]=-1.79$ defines the oldest and most metal poor limit of simple stellar populations by edging the oldest MSTO and the blue side of the RGB. By extending to the red edge of the RGB, outlined by the 7 Gyr old isochrone, this CMD region indicates a period of active star formation from $\sim13.5$ to $7.0$ Gyr ago. The isochrones for $\rm t=5.5$ Gyr, $\rm t=4.0$ Gyr, and $\rm [Fe/H]=-0.96$, which outline the brighter SGB and a corresponding part of the RGB, evidence the star formation activity at intermediate ages. It can be presumed from the plots that the region Fornax1 was likely forming stars at different epochs throughout its lifetime from 13.5 Gyr ago until as recently as a few hundred Myr ago.

Besides indicating the presence of stellar populations of different ages and metallicities, the good fit of the isochrones to the tip of the RGB and the RC confirms the accuracy of the magnitude shift adopted to account for the distance modulus and foreground extinction.

\section{SFH Procedure} \label{sec:sfh_procedure}

The preliminary analysis via comparison with isochrones presented in Section~\ref{sec:observed_cmd} suggests the diversity of stellar populations that comprise the studied regions. In this section we describe the quantitative analysis performed through the computation of precise SFHs with the CMD fitting technique. Our procedure followed closely the one described in detail in \cite{Aparicio2009} and \cite{Hidalgo2011}. Below is a brief overview of all the steps taken to obtain a SFH.

\subsection{Synthetic CMD} \label{sec:synth_cmd}

The first step of constructing a model CMD was to create a global synthetic population. The synthetic CMD presented here consists of $50\times10^6$ stars over a sufficiently large range of ages and metallicities, based on the observed ranges. The age and metallicity were limited to the intervals $\rm t = 0.03-13.9$ Gyr and $\rm -2.27 \leq [Fe/H] \leq -0.35$ (according to Figure~\ref{fig:obs_cmds_iso}).

The theoretical evolutionary framework needed to obtain the synthetic CMDs was based on the solar-scaled BaSTI stellar evolutionary library \citep{Pietrinferni2004}, previously adopted for construction of a synthetic CMD for the ASTs and the preliminary comparison between the observed CMDs and theoretical isochrones. The synthetic CMD was created following a constant (uniform) SFR in the age-metallicity plane defined within the quoted ranges. The IMF from \cite{Kroupa2002} was assumed with the fraction of the binary stars and their relative mass distribution $f=0.5$ and $q=0.4$, respectively. The latter were found to be the best combination from the studied ranges of these parameters: $f=[0.3,0.5,0.7]$ and $q=[0.1,0.4]$ and agreed with an estimate of these values in \citet[appendix therein]{Monelli2010a}. Mass loss along the RGB stage was set to $\eta=0.2$ \citep{Pietrinferni2004}.

Following the creation of the synthetic CMD, observational effects (see Section~\ref{sec:photometric_procedure}) were simulated in it using the code DisPar\footnote{Description of the code DisPar is available at: \url{https://riull.ull.es/xmlui/handle/915/3961}}, which applied the results of the AST for artificial stars (description of an alternative code called \textit{obsersin} is given in \citealp{Hidalgo2011}). It is important to account for these effects, as they scatter observed stars from their actual positions on the CMD, which is especially significant at faint magnitudes below the oldest MSTO \citep{Gallart1999}. The code investigated the colour-magnitude boxes of size $(col,mag)=(0.04,0.1)$ around a position of each synthetic star with the magnitudes $(V_s,I_s)$ and the same box in the space of the artificial stars with their input and recovered magnitudes $(V_i,I_i)$ and $(V_r,I_r)$, respectively. If the box contained at least 10 recovered artificial stars, one was selected randomly and its magnitudes were used to disperse the synthetic star as $V_s'=V_s-(V_i-V_r)$ and $I_s'=I_s-(I_i-I_r)$. If the artificial star was marked as unrecovered in the AST table, then the synthetic star was discarded. Otherwise, if a box had fewer than 10 AST stars, then a box of double size was used. A dispersed synthetic CMD is shown in Figure~\ref{fig:bundles}.

\subsection{Parametrising the CMDs} \label{sec:parametrising_cmds}

What we fit during the SFH procedure are distributions parametrised with colour and magnitude (in case of both the observed and synthetic CMD) and age and metallicity (in case of the synthetic CMD). The synthetic CMD can be divided in a number of simple stellar populations (SSP) defined by a finite range of age and metallicity. Arbitrary model CMDs can then be obtained by using a linear combination of the SSPs. The initial configuration of age and metallicity bins $t_k$ and $z_l$, respectively, was set to:

$t_k$ (Gyr) = [0.03, 0.1, 0.2, 0.3, 0.4, 0.5, 0.6, 0.7, 0.8, 0.9, 1.0, 1.2, 1.4, 1.7, 2.0, 2.5, 3.0, 3.5, 4.0, 4.5, 5.0, 6.0, 7.0, 8.0, 9.0, 10.0, 11.0, 12.0, 13.9],

$z_l$ = [0.0001, 0.00012, 0.00015, 0.00018, 0.00021, 0.00026,  0.00031, 0.00038, 0.00046, 0.00056, 0.00067, 0.00081, 0.00098, 0.00119, 0.00144, 0.00174, 0.00211, 0.00255, 0.00309, 0.00373, 0.00452, 0.00547, 0.00661, 0.008].

\begin{figure}
	\includegraphics[width=\columnwidth]{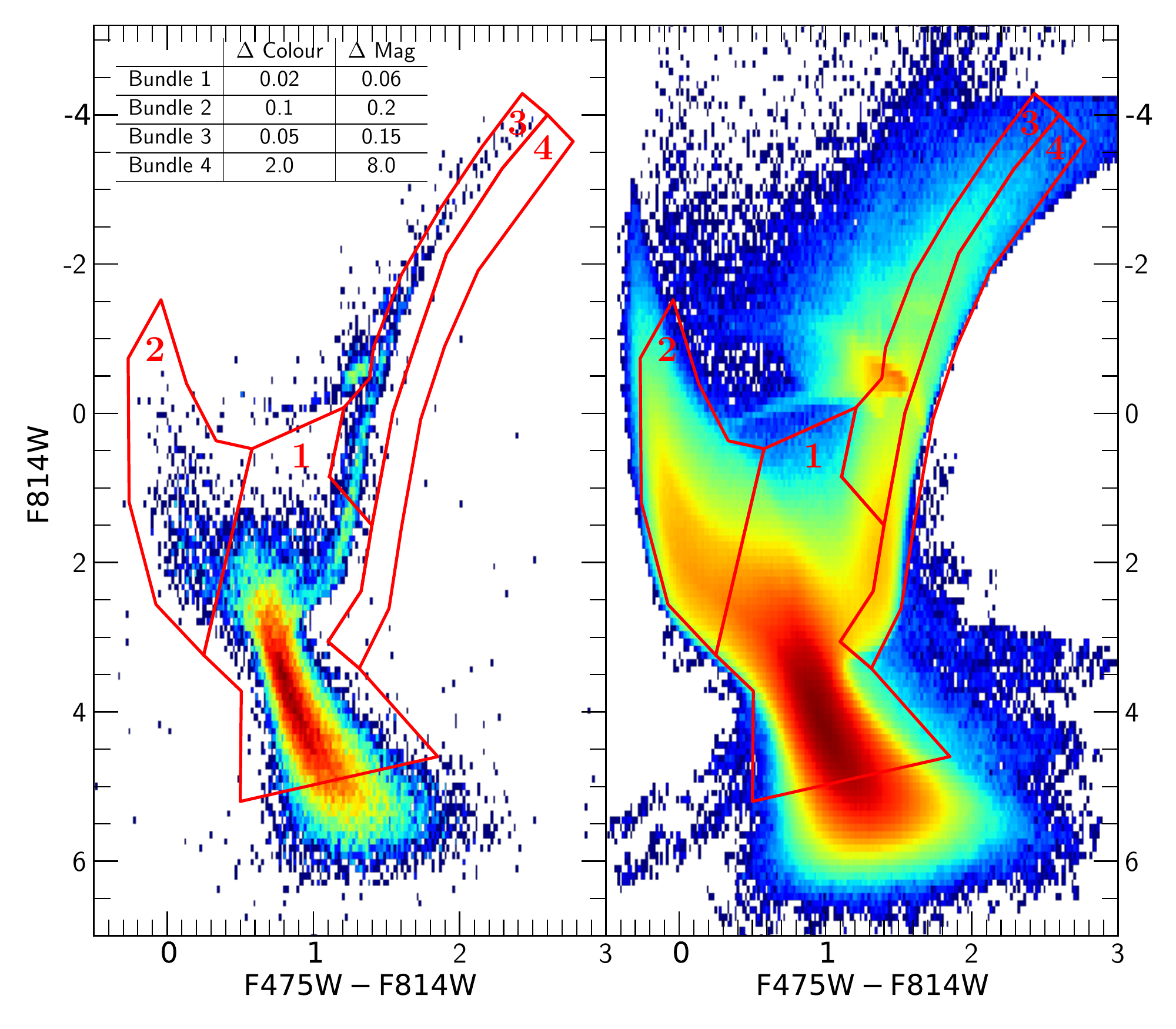}
    \caption{Observed CMD (Fornax1) and synthetic CMD (used for SFH calculations for both Fornax1 and Fornax2) after simulating observational effects. Bundles and bin sizes within them define the colour-magnitude parametrisation for TheStorm code (see Section~\ref{sec:parametrising_cmds}). The bundles are drawn around some of the key CMD regions, like MS and SGB (bundle 1), young MS (bundle 2), RGB (bundle 3). Bundle 4 is used to force the SFH code to avoid including stars where there are none observed. The respective sizes of grid boxes in colour and magnitude for all bundles are indicated in the table on the left panel.}
    \label{fig:bundles}
\end{figure}

In order to compare the distribution of stars in the observed and model CMDs, they were divided into ``bundles'', which outline CMD regions with different densities of stars. For example, separate bundles are usually selected to outline the heavily populated MS below the MSTO, young MS that has significantly fewer stars and intermediately dense RGB and RC. Further, the bundles were binned, where the aim was for every non-zero bin to contain a number of stars of the order $\sim$1-100, with at least $50\%$ of bins containing 5 stars to ensure the balance between the resolution of the SFH solutions and their signal-to-noise ratio. The adopted bundle design is illustrated in Figure~\ref{fig:bundles}, where the sizes of the bins in each bundle are indicated. It is worth noting that the RC was not included in the binning due to the disagreements between the RC observations and models described in detail in Section~\ref{sec:model_cmds}, which also describes the implications of choosing the bundle and bin configurations differently.

The number of stars $M_i^j$ distributed within every colour-magnitude box $j$ of the synthetic CMD was calculated along synthetic SSPs $i$ (with $i$ from $1$ to $n\times m$, where $n$ and $m$ are numbers of bins in age $t_k$ and metallicity $z_l$) and compared against the distribution of the observed stars $O^j$ in the corresponding box $j$ of the observed CMD. The procedure was aimed at reproducing the observed distribution $O^j$ with the model stars $M^j$ calculated by means of the linear relationship: $M^{j}=\sum_{i} \alpha_{i} M_{i}^{j}$, with $\alpha_i > 0$.

The best fit of the total distribution of model stars $M$ against $O$ for all $n\times m$ models was found by minimising the Cash statistic \citep{Cash1979}, treated as a goodness-of-fit measure:

\begin{equation} \label{eq:chi2}
    \chi^{2}=2\sum_{j} (M^{j}-O^{j}\ln{M}^{j}).
\end{equation}

It is analogous to the Gaussian ``chi-square'' quantity but derived from the Poisson log-likelihood to more accurately quantify the fit of the CMD bins with small numbers of stars. The reduced statistic $\chi^{2}_{\nu}=\chi^{2}/ \nu$ was used as a characteristic weight of the fit, where $\nu=b-n\times m$ is the number of degrees of freedom, $b$ is the number of colour-magnitude boxes defined in the CMD. The calculation of the optimal fit of the final model SFH and model CMD were performed using TheStorm code (see \citealp{Bernard2018}, which follows similar procedures as IAC-pop/MinnIAC used, for example, in \citealp{Aparicio2009}, \citealp{Monelli2010a,Monelli2010b}). The details of the algorithm procedure are provided in the following section.

\subsection{Solving for a model CMD} \label{sec:solving_model_cmd}

A solution for the SFH can be generally described as an array of the scaled alpha coefficients $\psi(t,z)=A\alpha(t,z)$ derived linearly, by estimating coefficients $\alpha_i$ for each SSP $i$, and scaled by the constant $A$, which is related to the stellar mass $M_{mod}$ of a model CMD, the area of the field of view S and age and metallicity range $\Delta t$ and $\Delta Z$, respectively, as $A=M_{mod}/(\Delta t\times \Delta z \times S)$. This array represents a 2D distribution $\alpha(t,z)$ with dimensions $[k, l]$ of age and metallicity, i.e. age-metallicity relation, which can be marginalised over all $l$ to obtain the SFR $\psi(t)$ at times $t_k$: $\psi_{k}(t)=A\sum_{l} \alpha_{k,l}(t,z)$. The metallicity enrichment $z(t)$ can be estimated as a median of all products of $z_l$ multiplied by normalised alpha coefficients $\alpha_{k,l}/\sum_{l}{\alpha_{k,l}}$ at all times $t_k$: $z_k(t)=A\prod_{l} median[\alpha_{k,l}(t,z)z_{l}]$.

\begin{figure}
	\includegraphics[width=\columnwidth]{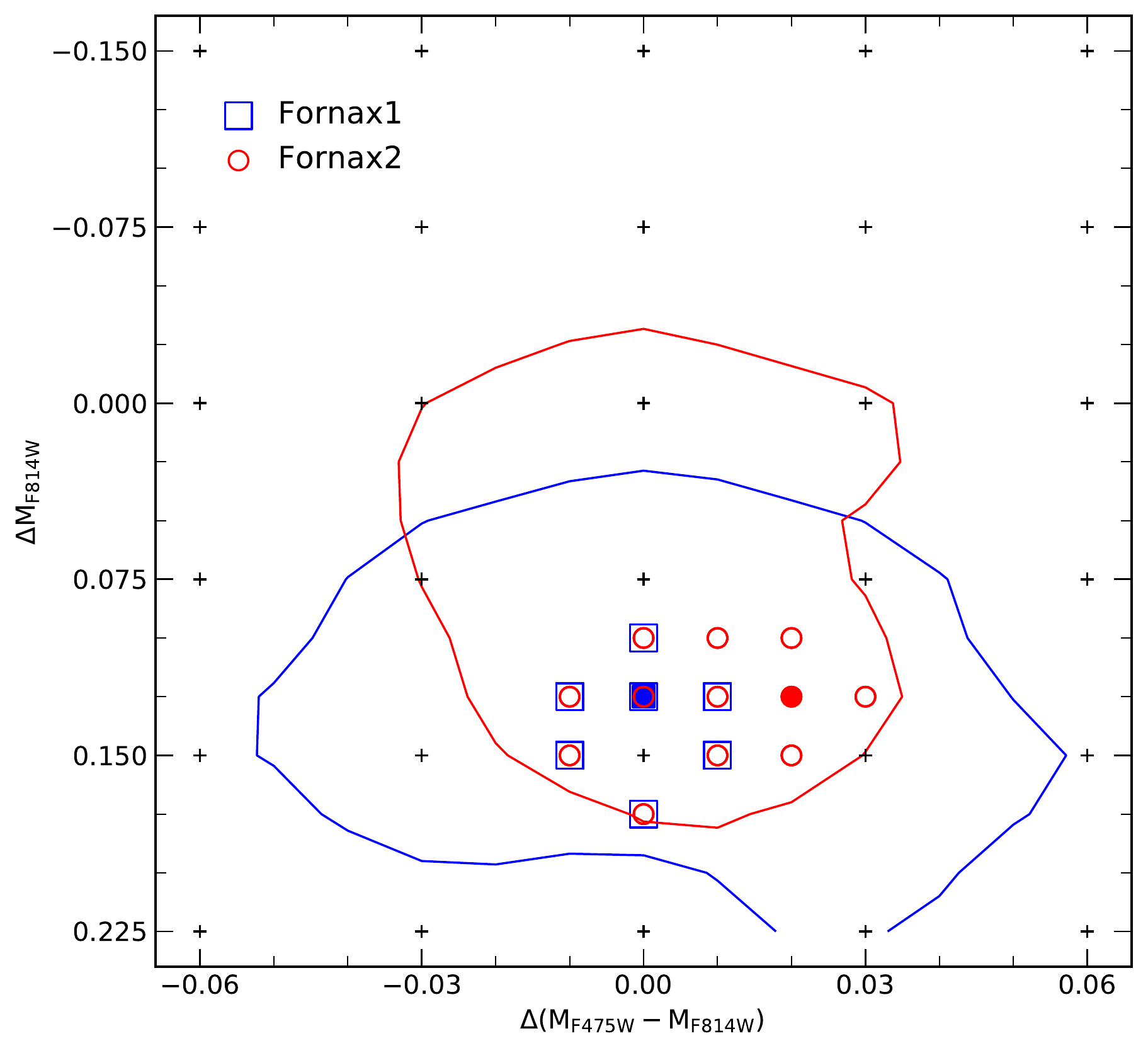}
    \caption{Likelihood map of colour and magnitude shifts sampled during the SFH procedure. The $+$ markers identified the common grid. The squares and circles represented the finer sampling grid for Fornax1 and Fornax2, respectively. The best positions in the respective fine sampling grids were marked with the filled square and circle. The blue and red contours indicate one-sigma regions for Fornax1 and Fornax2, calculated as the combination of the errors of the solutions with shifted bins at the best offset and solutions with the Poisson-resampled observed stars (see text for description). The likelihoods are smoothed with a Gaussian kernel (with 3-pixel window with $\sigma=3$ pixels).}
    \label{fig:chi2map}
\end{figure}

Distance modulus $\mu$ and colour excess $E(B-V)$ (Section~\ref{sec:observed_cmd}) were used to convert the observed CMD to absolute magnitudes. A grid around the adopted $\mu$ and $E(B-V)$ values was sampled by the code to account for the uncertainties in distance modulus, reddening and photometric zero points using the following colour and magnitude offsets: $\Delta (M_{F475}-M_{F814W})=[-0.06, -0.03, 0.0, 0.03, 0.06]$ and $\Delta M_{F814W}=[-0.15, -0.075, 0.0, 0.075, 0.15, 0.225]$, which resulted in 30 positions (marked by '+' markers in Figure~\ref{fig:chi2map}).

After the position of the minimum $\chi^2_{\nu}$ was identified, the code refined the search with smaller colour-magnitude shifts (to the left/right, top/bottom, totalling 4 additional shifts) around the position (shown as squares for Fornax1 and circles for Fornax2). The finer sampling was repeated around every newly identified minimum until convergence. At every offset, the calculation was repeated 5 times, each time applying small random shifts to the initial guess of the SFH solution, to avoid local minima, and the median of the 5 solutions was used to characterise each offset. Finally, this strategy resulted in 37 sampled offsets ($37 \times 5=185$ SFHs) for Fornax1 and 42 ($42 \times 5=210$ SFHs) for Fornax2, with the most likely positions marked by the filled square and circle for Fornax1 and Fornax2. The optimal colour and magnitude offsets were applied to the observed CMDs. As seen from the figure, both Fornax1 and Fornax2 CMDs were systematically forced towards fainter magnitudes for the best fit with the model CMDs, within the uncertainty in the distance modulus.

Once the position of the optimal offset was determined we estimated the uncertainty in the resulting solution using the following procedure. Given the parametrisation, no error estimate was associated with each SFH calculation, as the best-fit model CMD was computed by using a numerical iterative optimiser, which converged to the solution without sampling the whole involved parameter space. Instead, three main errors sources affecting the best solution were considered: the uncertainty in the selection of the colour, magnitude, age and metallicity bins, the Poisson uncertainty in the binning of the observed CMD and the uncertainty in the distance and reddening offset (as well as photometric zero points). This procedure followed the one described in \cite{Hidalgo2011}.

Regarding the first error source, the code introduced changes to the colour, magnitude, age and metallicity bin configurations: the bins were shifted by different fractions of their width, with the boundaries of their total ranges fixed. Overall, 12 different bin configurations were investigated, and an average of the 12 solutions weighted by their corresponding $\chi^{2}_{\nu}$ statistics was taken as the best model. This step was expected to account for the uncertainties in the human-selected parametrisation of the CMDs.

Then, Poisson resampling of the colour-magnitude distribution of the observed CMD was performed 6 times to recalculate the solution with its optimal parameters fixed. The standard deviation of the 6 resultant solutions was propagated to the error in the best solution.

Finally, the regions in the space of colour and magnitude offsets shown as blue and red contours in Figure~\ref{fig:chi2map} were used to propagate the corresponding error to the best SFH solution. In particular, likelihoods of solutions at every position in the map of offsets were calculated based on the employed reduced $\chi^2_{\nu}$ statistic:

\begin{equation} \label{eq:lik}
    L = e^{-\frac{\chi^2_{\nu}}{2}},
\end{equation}

and the combination in quadrature of the standard deviation of likelihood values of the solutions with the shifted bins (first error source, as above) and Poisson-resampled solutions (second error source) was used to calculate the blue and red confidence regions. Then, the standard deviation of all SFH solutions inside each of these confidence regions was calculated and propagated to the error of the corresponding best solution.

Additionally, an attempt to quantify the uncertainty associated with the SFH procedure and its ability to recover observed star formation bursts is presented in Section~\ref{sec:age_res_test}, where the age resolution of the procedure was estimated.

\begin{figure*}
	\includegraphics[width=\textwidth]{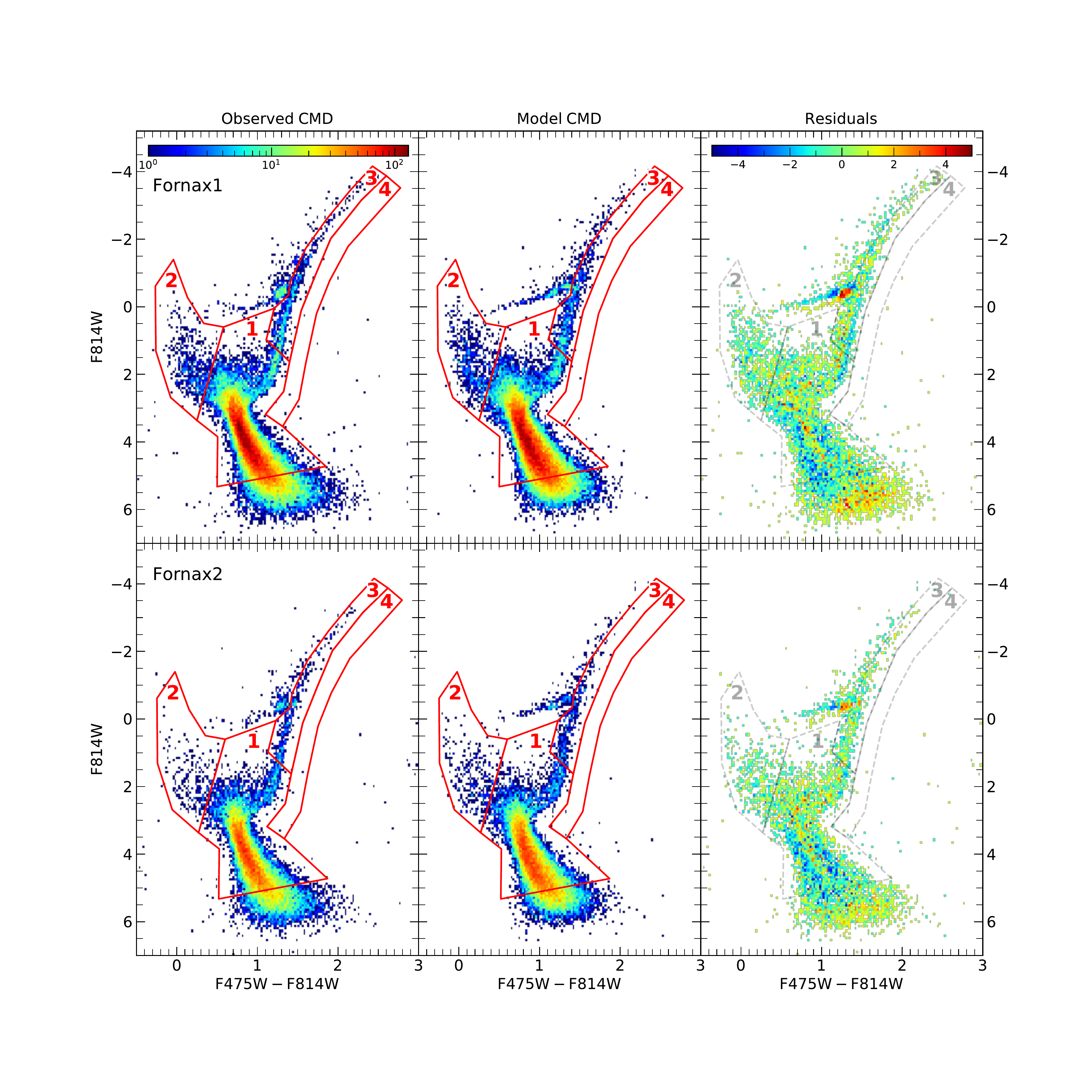}
    \caption{Left panel: observed CMDs of Fornax1 (top row) and Fornax2 (bottom row). Middle panel: final solution CMDs for the both regions. Right panel: residuals of subtraction of the model CMD from the observed CMD, defined as Poisson errors. Bundles 1-4 are the same as defined in Figure~\ref{fig:bundles}.}
    \label{fig:obs_sol_res}
\end{figure*}

\subsection{Solution CMDs of Fornax} \label{sec:model_cmds}

The final solution CMDs are presented in the middle panel of Figure~\ref{fig:obs_sol_res} compared to the observed CMDs in the left panel. The figure also shows the residuals $r_j$ based on Poisson statistics, defined as: 

\begin{align} \label{eq:residuals}
    r_j = \frac{O^{j} - M^{j}}{\sqrt{0.5(O^{j} + M^{j}) + 1}},
\end{align}

where $O^{j}$ and $M^{j}$, as before, are the numbers of stars in the $j$-th colour-magnitude bin of the observed and model CMDs, respectively. In general, the two solution CMDs demonstrate a good fit to the observations, with star counts in the CMD generally recovered within $\lesssim 2 \sigma$. The larger residuals in the lower MS could be caused by the low completeness in the photometry in this faint portion of the CMD, which has not been taken into account in the fit. Other strong residuals, over $2-3\sigma$, can be seen around the position of RC, HB and along the RGB.

The RC and HB poor fits have a known origin related to some physical prescriptions in the adopted stellar evolution models: the brightness of both RC and HB depends on the adopted conductive opacity setting, and the presently adopted version of the BaSTI library relies on - nowadays - outdated opacities (we refer to \citealt{cassisi:07} for a detailed discussion on this issue). The use of more updated conductive opacities (such as those provided by \citealt{cassisi:07} and used by \citealp{Hidalgo2018}) would have the effect of decreasing the brightness of RC and HB stellar models by $0.08-0.1$ magnitudes, which would contribute to improve the match between the synthetic and observed CMDs. As far as it concerns the color distribution along the RC/HB, it is mostly driven by the mass loss efficiency during the previous RGB stage. To properly model this feature of the synthetic CMD along the core He-burning sequence, one should introduce two free parameters in the CMD computation, which include the average mass loss efficiency $\eta$ and dispersion around this value, as was made for instance by \cite{savino:19}. On the other hand, the code adopted in the present investigation accounts for a fixed average mass loss efficiency during the RGB stage (i.e. $\eta=0.2$).

The origin of the $\sim 2 \sigma$ residuals in the RGB is less clear. Figure~\ref{fig:obs_sol_res} shows that model RGB is slightly steeper than the observed RGB (i.e., redder than the observed RGB below the RC and bluer at the brighter part). Since the RGB of metal poor populations are steeper than those of metal rich ones, this could indicate that a metallicity lower than the true metallicity is being recovered for the Fornax stellar population. However, the same trend in the residuals along the RGB can be observed in other SFH derivations using the same stellar evolution library \citep[e.g.][for the dSphs Cetus and Tucana and for the Large Magellanic Cloud (LMC)]{Monelli2010a, Monelli2010b, Meschin2014}, which could indicate some systematics in the RGB color in the models, most likely related to inaccuracies in the color transformations which have, for example, some dependency on the exact chemical composition adopted \citep{Pietrinferni2006}. In fact, in \citet{Meschin2014}, the modelled age-metallicity distribution was shown to be accurate via comparison with that derived from spectroscopic datasets, within uncertainty. In Appendix~\ref{sec:mdf} we perform a comparison between the metallicity distribution function (MDF) derived from the SFH of Fornax and that derived from spectroscopic measurements by \cite{Battaglia2006}, \cite{BattagliaStarkenburg2012} and \citet{Kirby2010}. It shows that indeed the former may be slightly more metal poor, especially in comparison with the MDF from \cite{Kirby2010}. However, note that they warn about a possible bias in their sample due to the fact that some stars on the extreme blue end of the RGB (where asymptotic giant branch, extremely young or extremely metal-poor red giant stars would be located) did not pass the spectroscopic selection criteria. This bias would lead to underestimating the fraction of metal poor stars in that spectroscopic sample.

\section{SFH of Fornax} \label{sec:sfh_results}

\begin{figure}
	\includegraphics[width=\columnwidth]{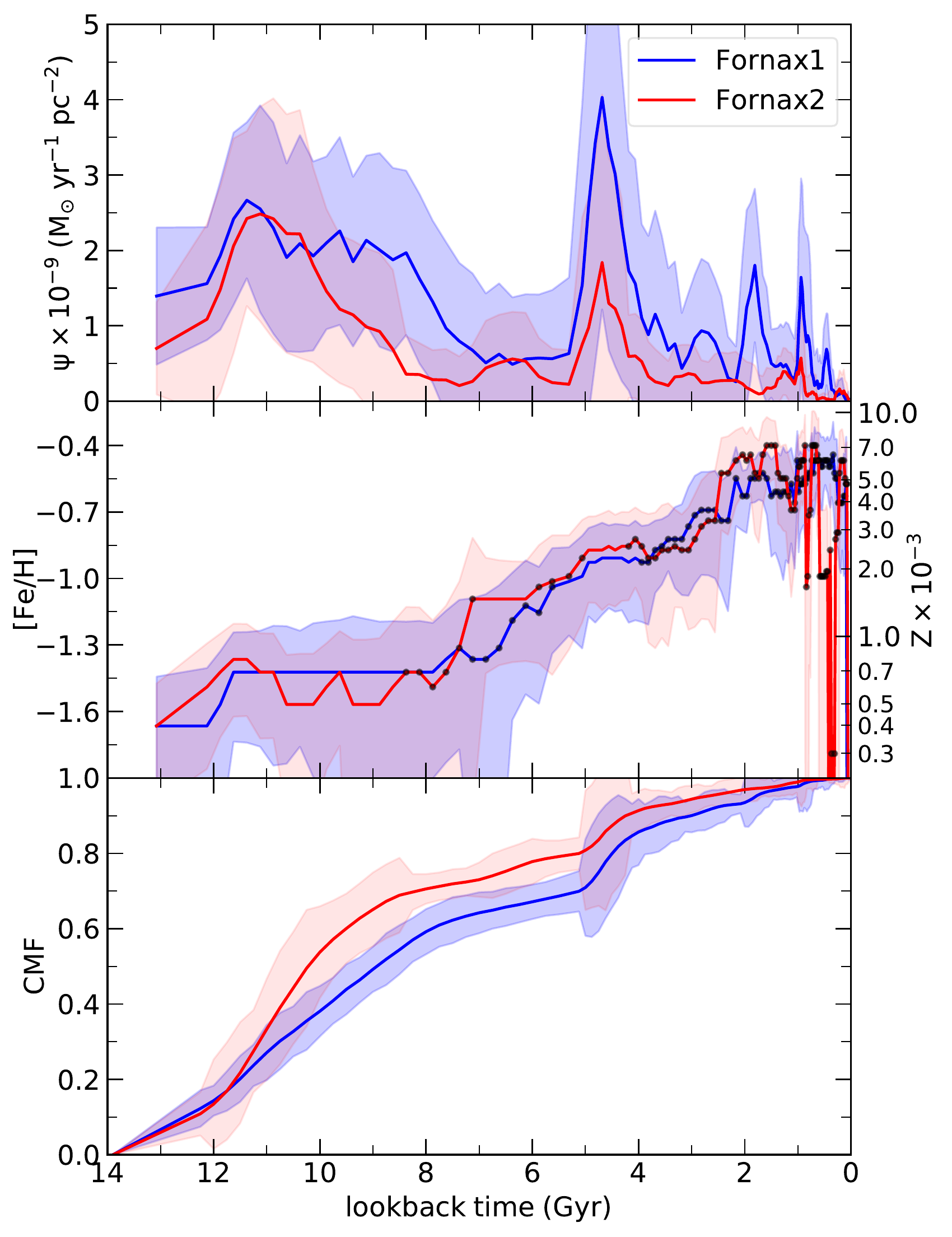}
    \caption{SFH of Fornax1 and Fornax2, shown by the functions of SFR $\psi(t)$ (top panel), chemical enrichment law (CEL) (middle panel) expressed in terms of the Fe abundance (left axis) and $z(t)$ (right axis), cumulative mass function (CMF) $\Psi(t)$ (bottom panel). Errors in the SFR and CEL were propagated from the combination of solutions with slightly different parametrisation, solutions with the Poisson-resampled observed stars and solutions with different distance and reddening offsets (see Section~\ref{sec:solving_model_cmd} for the description of the error sources). The CMF was calculated from the SFR with the error propagated correspondingly. The black dots in the middle panel indicate the age bins with the least statistical weight, containing $<1\%$ of total stellar mass.}
    \label{fig:sfhs}
\end{figure}

The individual solutions for the SFH of Fornax1 and Fornax2 are shown overplotted on Figure~\ref{fig:sfhs}. The blue and red shaded error bars are calculated as a combination of three error sources as described in Section~\ref{sec:solving_model_cmd}. Both regions have a similarly extended and non-uniform history of star formation. As indicated by the SFR (upper panels), the most active and extended period of star formation occurred in the oldest and most metal-poor epoch ($\rm -1.7 < [Fe/H] < -1.3$, from the AMR on the middle panels) in both regions.

The periods lasted until $\sim7$ and $\sim8.5$ Gyr ago, during which they formed almost 65 and 70\% of stellar mass in Fornax1 and Fornax2, respectively (as can be seen by the cumulative mass functions CMF at the bottom panels). In fact, as shown later in Section~\ref{sec:backgr_sfh}, the stars formed during $\sim5.5-7$ Gyr ago in Fornax1 and, possibly Fornax2, may belong to the oldest burst instead and appear shifted towards younger ages due to the observational effects. In such case the fractions of mass in stars formed in the first star formation episode could reach 70 and 80$\%$ in Fornax1 and Fornax2, respectively.

A prominent second peak in the SFR, centred at $\sim4.6$ Gyr ago, accounts for the $\sim5.1-4.1$ Gyr-old stars with $\rm -1.0 < [Fe/H] < -0.8$ in both Fornax1 and Fornax2. There is also a small bump in the SFR of Fornax2 at $6.4\pm0.6$ Gyr ago, which is of the same intensity as the low-level SFR in Fornax1 at the same lookback time.

The primary distinction between the two regions is the possible presence of different separate star formation episodes in the last 3 Gyr, with measured peaks at lookback times $2.8\pm0.3$, $1.8\pm0.2$, $0.9\pm0.1$, $0.5\pm0.1$ Gyr (Fornax1); and $1.1\pm0.3$, $0.2\pm0.1$ Gyr (Fornax2). While the SFR of Fornax1 seems to occurs primarily in separate bursts, star formation in Fornax2 appears to be more quiescent, with indications of at least one distinct burst, around 1 Gyr ago occurring in both regions.  However, note that, given the errors, most of the young peaks and low-level star formation are compatible with zero rate (more on low-level SFR in Section~\ref{sec:backgr_sfh}). Despite the somewhat different recent star formation activity, the chemical enrichment at the recent epoch is estimated to be very similar ($\rm -0.7 < [Fe/H] < -0.5$) in both regions. Black dots in the middle panel of Figure~\ref{fig:sfhs} indicate the age bins in which $<1\%$ of total stellar mass has formed. Clearly, the statistical significance of the populations younger than 1.5-2 Gyr old is low; therefore the ages and metallicities of the derived young stellar populations are highly uncertain.

Additionally, the overall level of star formation is higher in Fornax1, with only the oldest burst being similar in magnitude in the two regions. The two solutions have a very similar overall trend in metal enrichment during the whole lifetime of the galaxy. The enrichment proceeded monotonically up to the current epoch in the range $\rm-1.6 < [Fe/H] < -0.4$.

A very interesting feature that can be clearly seen from the overlapping SFRs and CMFs on Figure~\ref{fig:sfhs} is that the old star formation epoch was more extended in time in the inner region, and later ceased in both. Then, similarly, the young star formation episodes lasted till more recently in the central field of Fornax1, compared to the Fornax2 field. This kind of spatial stellar population gradient, with central regions being younger on average, is commonly seen in dwarf galaxies \citep{Harbeck2001}. What is very remarkable in the case of Fornax is that this behaviour is observed twice, first in the old star formation epoch, and then at recent times. In either case, this behaviour may be attributed to different timescales of gas depletion in the inner and outermost regions of the galaxy.

As a test of the recovered SFH, solutions with different age and metallicity bins were trialed, although the adopted configuration provided the optimal balance between the age and metallicity resolution and uncertainties. For instance, increasing the bin sizes by $1.5-2.0$ times across the whole history of Fornax recovered the major bursts at similar ages compared with the solutions presented above but the stellar bursts were smoothed out decreasing the resolution of simple stellar populations. This, for example, led to broadening of the oldest and intermediate-age bursts lowering their SFR magnitude and introducing some background SFR at the intermediate epoch during $\sim5.5-8$ in both regions. Therefore, the initially adopted bin configuration was preferred for the higher resolution of the stellar populations in the former case. Further narrowing of the bins was not adopted, as it would decrease the signal-to-noise due to the small number statistics.

The implications of choosing different bundle strategies were illustrated in the study of the SFH of Leo A in \cite{RuizLara2018}. It was shown that although alternative bundle designs have a minor effect on the derived SFH of the galaxy, with most of differences included in the uncertainty, the bundle systems with the largest coverage of the observed CMD recovered best the metallicity distribution of the RGB stars derived from spectroscopy. However, here the bundle including the RC region was omitted as it resulted in higher residuals at the faint red part of the RGB and MSTO, while still not fitting the RC and HB accurately due to the issues with the stellar evolutionary models discussed above.

\subsection{Age resolution test} \label{sec:age_res_test}

In this section we present some tests aimed at (1) verifying the accuracy of the age of $\psi(t)$ episodes in our solutions (hereafter called $\psi_s(t)$) recovered with the BaSTI stellar evolutionary models, (2) testing the time resolution of SFR at different lookback times, given the uncertainties in the SFH derived from the photometric errors, the small number of stars in the observed CMDs, and the intrinsic limitation of the SFH recovery method. The tests were performed by reproducing the positions and widths of the $\psi_s(t)$ bursts with mock bursts created with synthetic stellar populations at a constant assumed SFR. Some examples of the test in previous works include \cite{delPino2013}, \cite{deBoer2012}, \cite{Hidalgo2011}, \cite{Aparicio2009}.

The mock populations representing the bursts were created with different age ranges and their corresponding metallicities (as per the obtained AMR for Fornax1 and Fornax2 in Figure~\ref{fig:sfhs}) and a constant input SFR $\psi_i(t)$. They were extracted from the synthetic CMD used for the SFH derivation (Section~\ref{sec:synth_cmd}) with a number of stars such that the recovered SFH closely reproduced the height of the $\psi_s(t)$ peaks. The mock CMDs for each burst were dispersed individually, according to the observational effects in the studied Fornax regions by using the results of the ASTs. The recovered (output) SFR of the mocks, $\psi_o(t)$, was computed by using the same procedure (with the same CMD parametrisation), as in Section~\ref{sec:parametrising_cmds}. Thus, by eliminating the uncertainties associated with matching the stellar evolutionary models with observations, distance and reddening estimates and keeping the observational errors it was possible to test the ability of the SFH code to recover the input $\psi_i(t)$ and test the age resolution of the recovered $\psi_o(t)$, ``smoothed'' by the simulated observational effects.

\begin{figure*}
	\includegraphics[width=\textwidth]{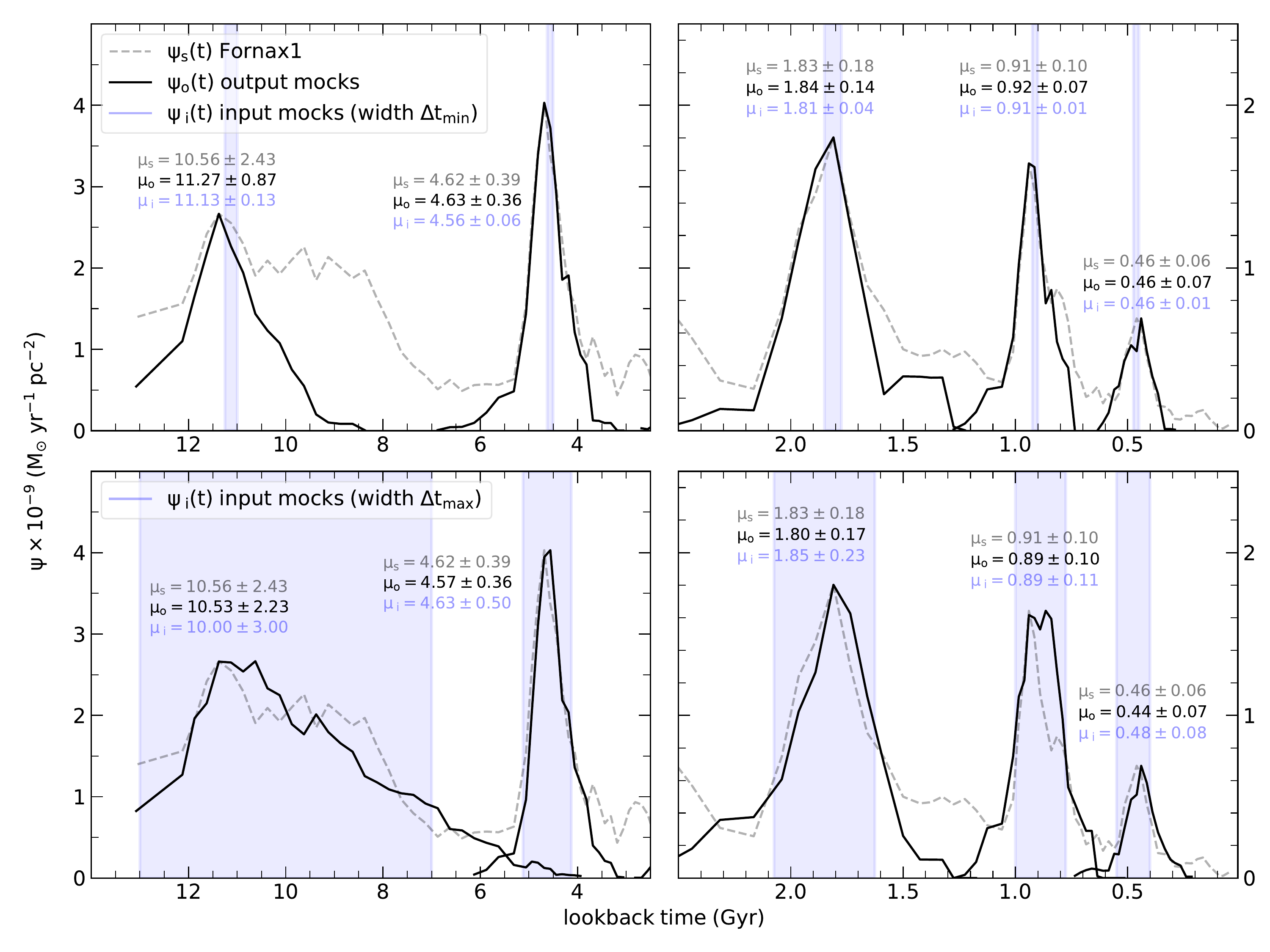}
    \caption{SFR recovered from the tests with the mock synthetic populations in Fornax1. Top panels show the tests with the smallest possible age widths of the input mocks, set by the sizes of the age bins. Bottom panels include the tests with the input mocks with the maximum width in age, such that the bursts could still be recovered. The age range of the input $\psi(t)_i$ is shown by the blue shaded areas and the magnitude of the shaded regions is not indicative of the $\psi(t)_i$ magnitude. The recovered mock synthetic bursts of $\psi(t)_o$ are shown in black solid lines, where the peak heights are normalised to the heights of the Fornax1 solution peaks. The solution $\psi(t)_s$ is shown in a grey dashed line. Peak mode positions and their standard deviations are shown as $\mu \pm \sigma$.}
    \label{fig:mocks1}
\end{figure*}

\begin{figure*}
	\includegraphics[width=\textwidth]{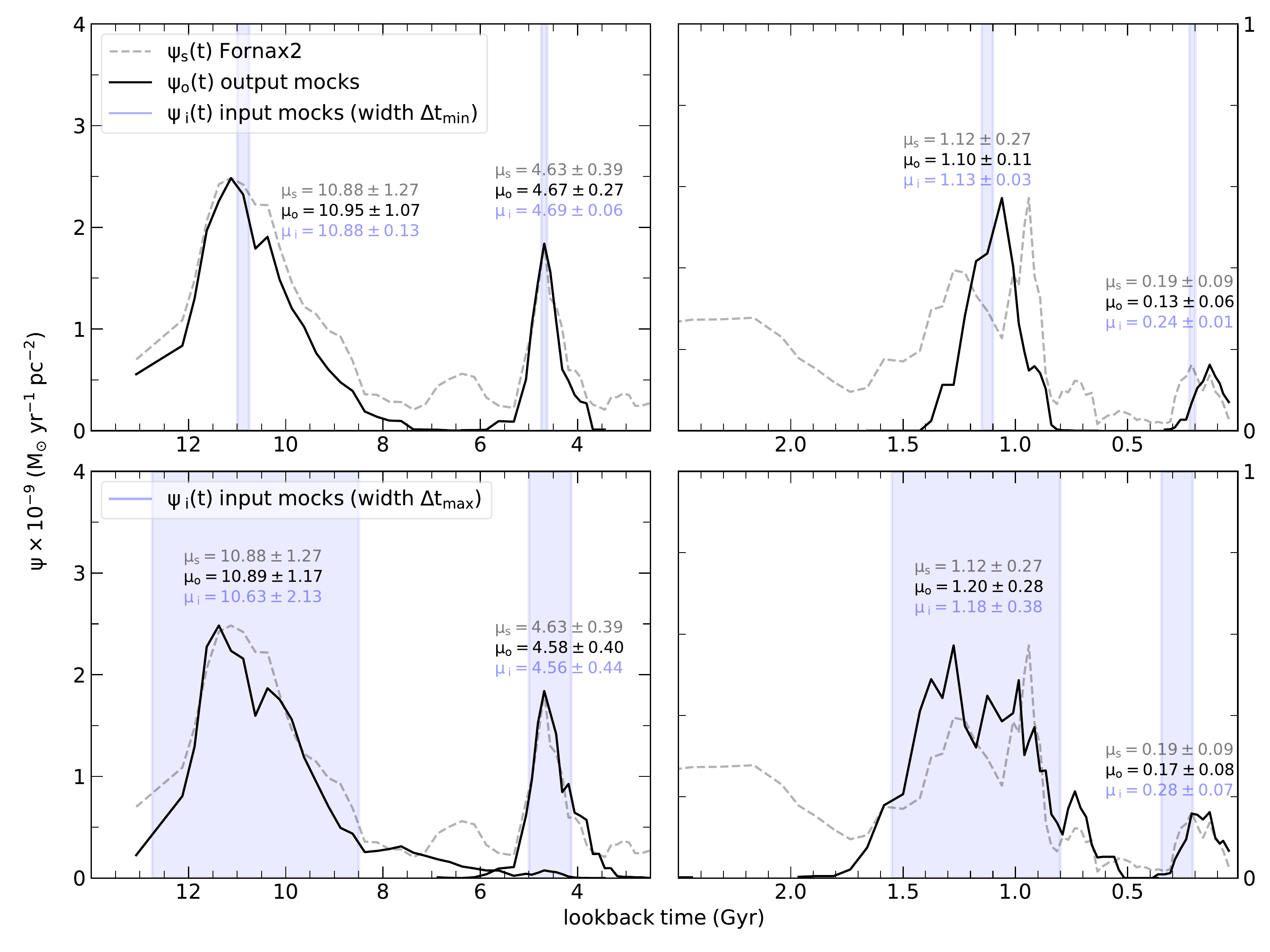}
    \caption{The same as Figure~\ref{fig:mocks1}, but for Fornax2.}
    \label{fig:mocks2}
\end{figure*}

The central ages of $\psi_i(t)$ bursts were taken to coincide with the mode age of the $\psi_s(t)$ bursts at different lookback times. We performed the test for star bursts in Fornax1 and Fornax2 and show the results in Figures~\ref{fig:mocks1} and \ref{fig:mocks2}, respectively. The panels show the solutions $\psi_s(t)$ in grey dashed lines, input $\psi_i(t)$ as blue shaded regions (height of the shaded regions is not indicative of the $\psi_i(t)$ SFR magnitude, but rather shows its age range) and output $\psi_{o}(t)$ in black lines. Most peaks of $\psi_o(t)$ and $\psi_s(t)$ can be fitted well with a Gaussian profile. The age modes $\mu$ of the solution and output bursts and their standard deviation $\sigma$ are indicated on the plots with the corresponding subscripts. In the case of the input $\psi_i(t)$, the central age and half age width is listed.  The recovered bursts were normalised to the height of the corresponding solution bursts in grey. The top rows of the figures show the tests with minimum input age ranges $\rm \Delta t_{min}$, equal to the widths of corresponding single age bins, and the bottom rows illustrate the tests with the maximum $\rm  \Delta t_{max}$, at which the recovered mock bursts of $\psi_o(t)$ just started to become wider than the solution bursts. The experiment with the $\rm \Delta t_{min}$ input limit was used to estimate the age resolution of the mock solutions, which was calculated as the FWHM of the $\psi_o(t)$ bursts. The values of $\rm \Delta t_{max}$ represented the upper boundary on the duration of input star formation bursts $\psi_i(t)$, i.e. the true mock bursts. The values $\rm \Delta t_{max}$ were obtained by incrementally increasing the width of the input bursts. Therefore, the limits $[\rm \Delta t_{min}, \rm \Delta t_{max}]$ for every burst represent constraints of its true duration. The results for the duration and metallicities for each input mock burst can be found in Table~\ref{tab:mocks_data}.

\begin{table*}
 \caption{Tests with mock bursts in Fornax1 and Fornax2 (column 1). Ages of the input mocks were centred on lookback ages $\rm \mu_{i,min}$ (column 2) and $\rm \mu_{i,max}$ (column 6) with their widths $\rm \Delta t_{i,min}$ (column 3) and $\rm \Delta t_{i,max}$ (column 7), respectively, where the minimum widths were defined by the age bins of the solution (i.e. at the limit of age resolution) and the maximum limit was such that the peak could still be resolved, before becoming wider than the solution burst. The recovered widths of the mock populations ($\rm FWHM$ $(\psi_o(t))$) are stated as $\rm \Delta t_{o,min}$ (column 4), defined as the age resolution, and $\rm \Delta t_{o,max}$ (column 8), defined as the upper boundary on the burst duration. Each input mock burst has the associated metallicity range $\Delta z\rm _{i,min}$ (column 5) and $\Delta z\rm _{i,max}$ (column 9).}
 \label{tab:mocks_data}
 \begin{tabular}{c|cccc|cccc}
  \hline
  solution & $\rm \mu_{i,min}$ (Gyr) & $\rm \Delta t_{i,min}$ (Gyr) & $\rm \Delta t_{o,min}$ (Gyr) & $\Delta z\rm _{min}$ & $\rm \mu_{i,max}$ (Gyr) & $\rm \Delta t_{i,max}$ (Gyr) & $\rm \Delta t_{o,max}$ (Gyr) & $\Delta z\rm _{max}$\\
  \hline
  Fornax1  & 0.46 & 0.03 & 0.16 & 0.00620 - 0.00660 & 0.48  & 0.15 & 0.15 & 0.00578 - 0.00661 \\
        & 0.91 & 0.03 & 0.18 & 0.00478 - 0.00515 & 0.89 & 0.23 & 0.24 & 0.00452 - 0.00578 \\
        & 1.81 & 0.08 & 0.34 & 0.00515 - 0.00547 & 1.85   & 0.45 & 0.40 & 0.00426 - 0.00578 \\
        & 4.56 & 0.13 & 0.85 & 0.00223 - 0.00240 & 4.63  & 1.00 & 0.84 & 0.00184 - 0.00240 \\
        & 11.13 & 0.25 & 2.05 & 0.00067 - 0.00071 & 10.00    & 6.00 & 5.26 & 0.00010 - 0.00137 \\
  Fornax2 & 0.24 & 0.03 & 0.15 & 0.00547 - 0.00578 & 0.28   & 0.14 & 0.18 & 0.00550 - 0.00661 \\
        & 1.13  & 0.05 & 0.26 & 0.00373 - 0.00395 & 1.18  & 0.75 & 0.65 & 0.00500 - 0.00700 \\
        & 4.69 & 0.13 & 0.65 & 0.00240 - 0.00255 & 4.56 & 0.88 & 0.94 & 0.00240 - 0.00270 \\
        & 10.88 & 0.25 & 2.51 & 0.00067 - 0.00071 & 10.63 & 4.25 & 2.76 & 0.00010 - 0.00081 \\
  \hline
 \end{tabular}
\end{table*}

As a result of the tests, output median ages $\mu_o$ of mock bursts were recovered close to their input positions $\mu_i$, except for the oldest bursts, which were shifted towards older ages by a fraction of the standard deviation of the output burst $\sim0.2\sigma_o$ in Fornax1 and $\sim0.1-0.2\sigma_o$ in Fornax2, and the youngest burst in Fornax2, which was projected towards younger age from $\mu_i$ by $\sim 1-2\sigma_o$. These results showed that the ages of the real star formation activities at intermediate epoch were recovered accurately in the solutions, while the computed SFR of the oldest and youngest bursts may be skewed to older and younger ages from the actual bursts, respectively. The systematic effect could be caused by the observational effects or, in case of the young SFR burst, could be due to the low number of stars associated with it ($\sim300$ stars).

As can be seen from the figures, some of the shortest mock bursts ($\rm \Delta t_{min}$) were recovered with the same $\sigma_o$ as the mock bursts with the maximum input width ($\rm \Delta t_{max}$). Therefore, the true lower limit could not be established for those bursts (i.e. such bursts were not fully resolved and could have a shorter true duration). For example, the peaks of $\psi_s(t)$ at $\mu=0.46, 0.91, 1.83, 4.62$ Gyr ago in Fornax1 could be caused as well by real star formation lasting shorter than the duration of the corresponding input mocks of $\rm \Delta t_{min}=0.025, 0.025, 0.075, 0.125$ Gyr, but our SFH code would not be able to resolve them, given the size of our age bins. Nonetheless, we could claim the upper duration limit for all bursts. It is important to point out that the shape of the real star formation rate bursts would be more complicated than our assumed constant $\psi_i(t)$ and, hence, the above estimates of burst duration could serve only as the first-order approximations.

Finally, based on the narrowest recovered peaks, the age resolution of our SFR, defined as $\rm \Delta t_{o,min}=FWHM(\psi_o(t))$ (see Table~\ref{tab:mocks_data}), was found to deteriorate as $\sim$0.2, 0.2, 0.3, 0.9, 2 Gyr at the lookback times $\sim$0.5, 0.9, 1.8, 4.6, 11.3 Gyr for Fornax1; and 0.2, 0.3, 0.7, 2.5 Gyr at the ages 0.1, 1.1, 4.7, 11 for Fornax2.

\subsection{Intermediate star formation epochs} \label{sec:backgr_sfh}

As seen from the upper panel of Figure~\ref{fig:sfhs}, the complex SFH of Fornax is characterised by distinct bursts, separated by the epochs of a low-level SFR. Even though the error bars at these epochs are compatible with a null SFR, we further investigated our solutions to find out whether the low-level SFR is real or a result of stars being ``dispersed away'' from their main star formation epoch by the observational effects in the process of SFH derivation.

\begin{figure}
	\includegraphics[width=\columnwidth]{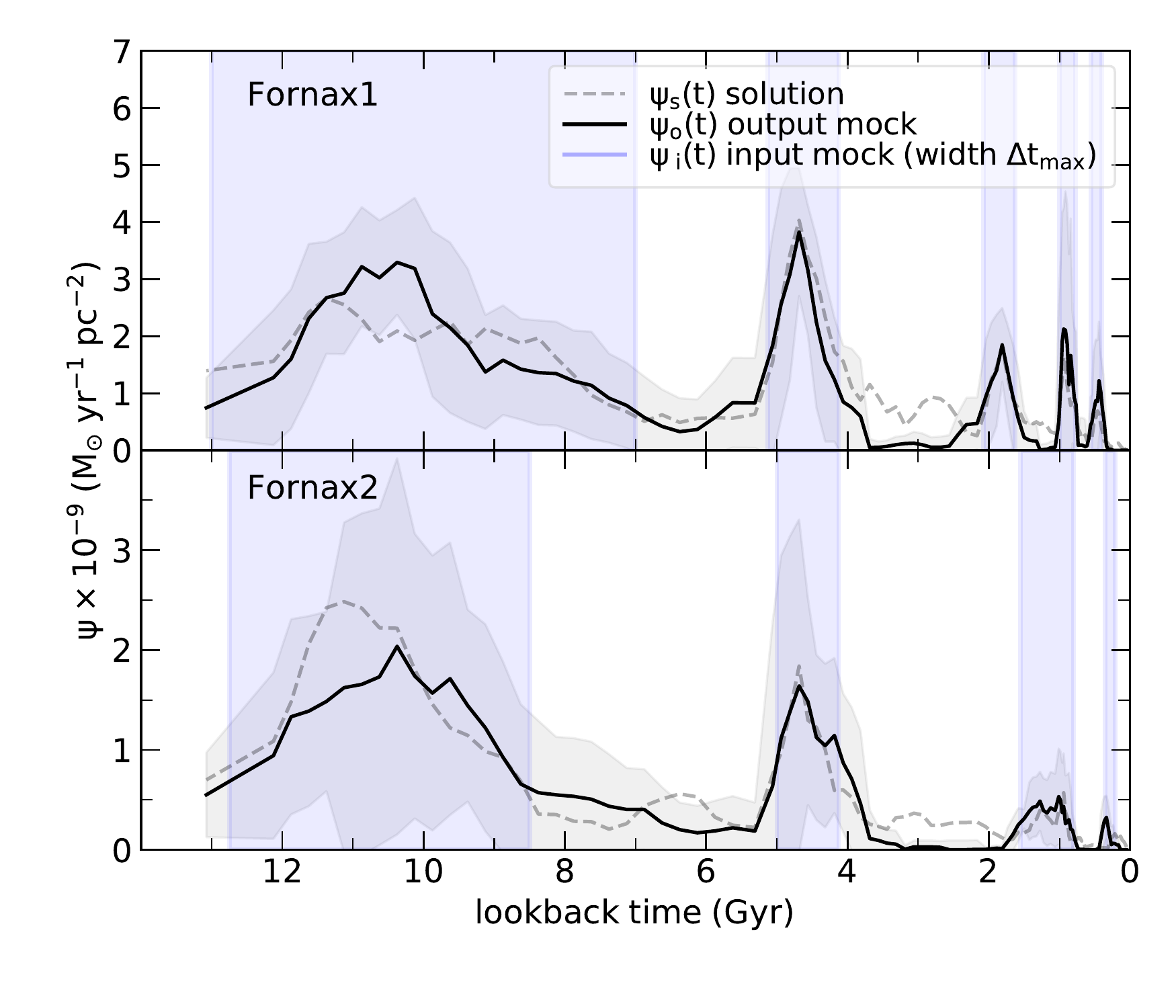}
    \caption{An attempt to reproduce the solution SFR of Fornax1 (top panel) and Fornax2 (bottom panel) by using only the synthetic bursts corresponding to the major star formation episodes (age range of these bursts is indicated by the blue shaded regions) and no low-level SFR between them. The input mocks had the maximum estimated duration $\rm \Delta t_{max}$ (see Table~\ref{tab:mocks_data}). The grey shaded area represents the error of the recovered $\psi_o(t)$.}
    \label{fig:sfr_mocks}
\end{figure}

To test this low level SFR we did an additional experiment with mock stellar populations that represented the main measured star formation bursts, not including the low-level star formation between them. For this purpose, we adopted the mock bursts from the age resolution test (Section~\ref{sec:age_res_test}). The input duration of the bursts was set to the maximum values $\rm \Delta t_{max}$ from Table~\ref{tab:mocks_data}, so as to have as many stars ``dispersed away'' by the observational effects as possible. In contrast to the age resolution test, the CMDs for all mock bursts were combined into a single CMD, from which the SFH was calculated by following the standard procedure (Section~\ref{sec:sfh_procedure}).

Figure~\ref{fig:sfr_mocks} shows that the mock bursts alone can account for the the $\psi_s(t)$ activity at $\sim8-5.5$ Gyr ago, between the oldest and intermediate star formation epochs, in both Fornax1 and Fornax2; however, when setting the width of the burst at $\sim$4.6 Gyr ago to $\rm \Delta t_{min}$ (not shown on the figure), the same background activity is still recovered with the mocks in Fornax1, but not in the case of Fornax2, suggesting that Fornax2 could have experienced star formation $\sim8-5.5$ Gyr ago if the duration of the burst $\sim$4.6 Gyr ago was close to $\rm \Delta t_{min}$. Also, interestingly, in the case of $\rm \Delta t_{max}$, the observed background SFR was not recovered in the period $\sim3.75-2$ Gyr ago in Fornax1 and Fornax2 and in between the youngest bursts in Fornax1, indicating the possibility of active low-level star formation at that time as well.

In conclusion, the test suggests that these regions may have experienced active star formation at the periods indicated by our SFH solutions (Figure~\ref{fig:sfhs}), including low-level star formation at some time $\sim8-5.5$ Gyr ago in Fornax2, $\sim3.75-2$ Gyr ago in Fornax1 and Fornax2 and at even younger ages in Fornax1. However, a halt in the star formation activity after the first burst may have occurred in Fornax1 and possibly also in Fornax2.

\subsection{On the impact of a different stellar evolutionary framework on the SFH retrieval}

Many SFH studies of LG dwarfs, including the previous SFHs of Fornax, have been conducted using the BaSTI models, thus relying on the same theoretical framework in this study was required for proper comparison between the findings in Fornax and other LG galaxies.

The implications of adopting different models are worth noting. For example, the updated version\footnote{The updated BaSTI library is available at: \url{http://basti-iac.oa-abruzzo.inaf.it/index.html}} of the BaSTI library \citep{Hidalgo2018}, which incorporates atomic diffusion, more updated physical inputs such as new conductive opacity tabulations and a different mass loss efficiency during the RGB stage, is expected to provide more accurate and reliable stellar models for the advanced evolutionary stages with respect to the models used presently \citep{Pietrinferni2004}, while producing the same results for the H-burning stage, as discussed in detail in Section~\ref{sec:model_cmds}. A detailed comparison of the updated version with the one used here and other recent libraries is presented in \citet{Hidalgo2018}. The efficiency of atomic diffusion has an effect on the old age dating: in general, for a given physical framework, in passing from models with no diffusion to fully efficient diffusion, the estimated ages decrease by 0.7-0.8 Gyr at old ages (i.e. ages larger than about 8~Gyr). Different libraries are expected to produce somewhat different results for the intermediate- and young-age populations depending on the adopted efficiency for the core convective overshooting during the H-burning stage, among other parameters (see \citealp{Gallart2005} for a general overview in the context of the SFH procedure). Some works have explicitly compared SFH results obtained adopting different star evolution libraries. For example, \citet{Monelli2010a} and \citet{Monelli2010b} did a comparison of SFHs at old ages obtained with the \citet{Girardi2000} and BaSTI libraries finding that the former produced slightly younger ($\lesssim 1$ Gyr solutions at old age; \citet{dolphin2012} performed a comparison between Padova \citep{Girardi2000}, BaSTI and Darmouth \citep{Dotter2008} libraries for both old and intermediate age bursts using synthetic populations.

In the case of the stellar population analysed in this work, we estimate that, if models with different convective overshooting were adopted, the main star formation bursts found to occur $\sim4.6$ and $\sim10.6$ Gyr ago would likely not be affected, while the bursts estimated to occur in the past 2 Gyr could be recovered at an older age depending on the adopted overshoot prescription. However, given the low statistical weight of star formation bursts in the recent $\sim 2$ Gyr, this would not change the main conclusions of the work. If atomic diffusion was assumed to be fully efficient (the diffusion is not implemented in BaSTI library used in this work), the two oldest, main star formation bursts could be shifted by up to $-0.8$ Gyr towards the current epoch. Given the age resolution test (see Section~\ref{sec:age_res_test}), such extreme change would still be within the uncertainty of the solutions for the oldest burst in Fornax1 and Fornax2, but just outside the error margin of the bursts at $4.6 \pm 0.5$ Gyr ago and $4.6 \pm 0.4$ Gyr ago in Fornax1 and Fornax2, respectively.

\subsection{Comparison with previous results} \label{sec:comparison}

In this section, we compare our SFH with previously published determinations from CMD analysis and with chemical evolution scenarios inferred from spectroscopy of individual stars.

\subsubsection{Star formation history} \label{sec:comparison_a}

The photometry and SFH results presented in this paper for Fornax1 and Fornax2 are widely consistent with the current view of Fornax as a galaxy with a complex SFH continuing to almost the present time. However, the extremely deep and precise CMDs resulting from the high signal-to-noise and high resolution HST images allowed us to provide important new details about the SFH by accurately dating and constraining the duration of the main SFR bursts. This information allows us to characterise Fornax as a galaxy that has experienced distinct SFH bursts separated by intermediate periods of null or low-level activity.

The analysis of two fields at different galactocentric distances allowed us to see the features of the stellar population gradient, reported by every wide-field study of stellar populations of Fornax starting from \cite{Stetson1998} and followed by \cite{Battaglia2006}, \cite{deBoer2012}, \cite{deBoer2013}, \cite{delPino2013}, \cite{delPino2015}, \cite{Bate2015}, \cite{Wang2019b}. In general, the oldest stars are found to be distributed quite uniformly within the tidal radius of the galaxy and the intermediate- and young-age stars are represented by a larger fraction in the centre.

The study by \cite{Buonanno1999} argued that Fornax experienced a second major burst of star formation at $\sim4$ Gyr ago, as judged by the separate SGBs in their CMD. This has been confirmed by the recovered peaks of $\psi(t)$ in our solutions. The age of the narrow episode at $4.6\pm0.4$ Gyr ago agreed very well in multiple averaged solutions for Fornax1 and Fornax2. However, this feature of the Fornax SFH, suggested long ago by the qualitative analysis of a high-quality WFPC2/HST CMD in the latter study, was not found in the more recent quantitative SFH determinations \citep{deBoer2012, delPino2013, Weisz2014}, while indications of it were reported by \citet{Coleman2008}. 

\begin{figure}
	\includegraphics[width=\columnwidth]{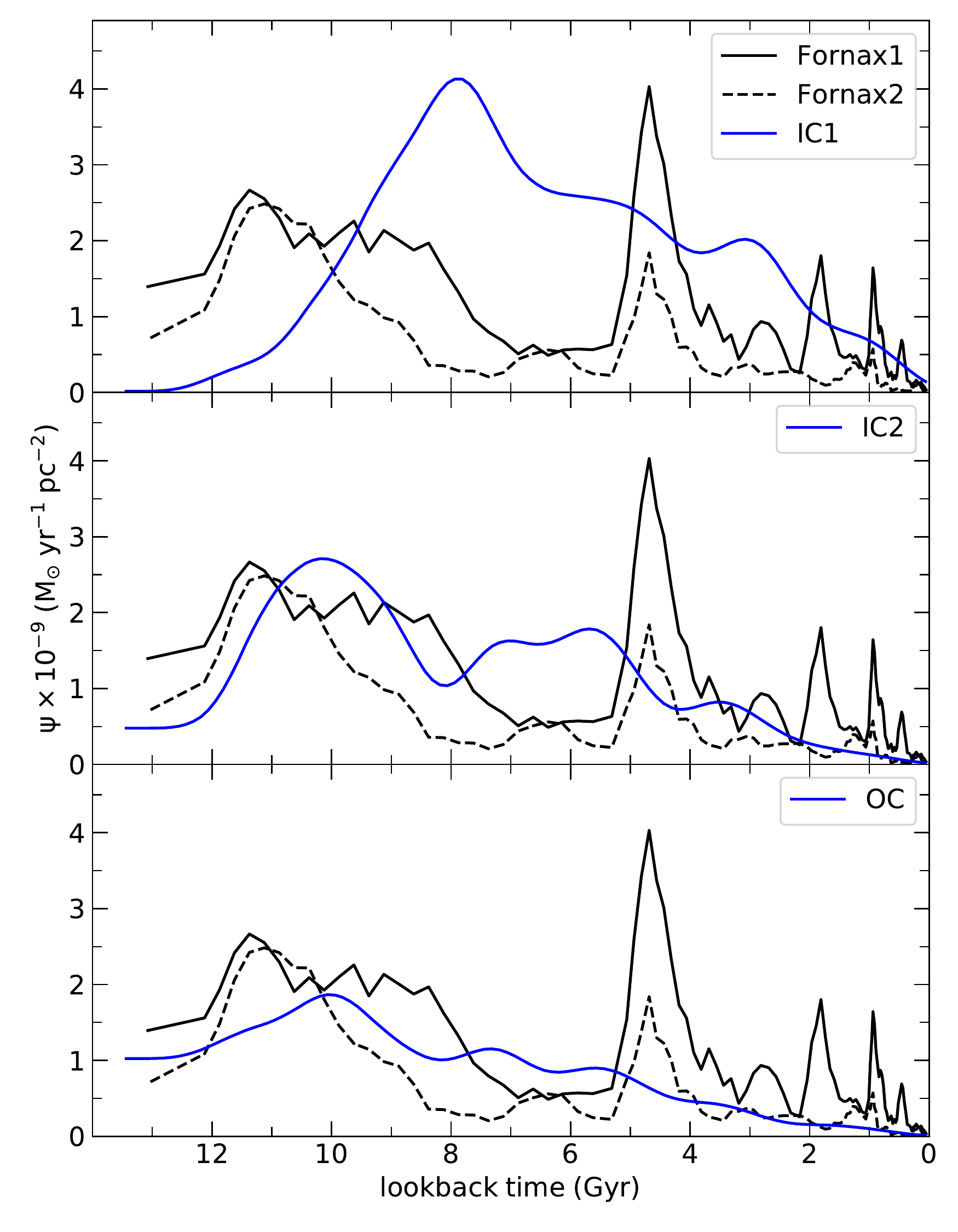}
    \caption{Comparison of the SFR functions obtained for regions Fornax1 and Fornax2 with IC1, IC2 and OC from \citet{delPino2013}.}
    \label{fig:sfhs_del_pino}
\end{figure}

The work by \citet{delPino2013} presents the first Fornax SFH derived from CMDs that reach the oMSTO in a range of galactocentric distances. Their SFH is superimposed on ours in Figure~\ref{fig:sfhs_del_pino} and the map of the corresponding field locations is shown in Figure~\ref{fig:footprints}. The SFR in the regions IC2, OC and, especially, the most crowded region IC1, are very smooth and do not show signs of the distinct star formation bursts found in our solutions. The most striking disagreement is found between the SFH of Fornax1 and IC1, {which are, in fact, the most directly comparable regions at all ages considering the asymmetry of the spatial distribution of the stellar populations younger than $\sim$2 Gyr, as shown in \cite{delPino2015}}. The SFH from \citet{delPino2013} demonstrates a $\psi(t)$ peak $\sim8$ Gyr ago that lasted for the whole intermediate-age epoch, while the active period in Fornax1 stopped after $\sim7$ Gyr ago and, likely, no stars were formed until $\sim5.1$ Gyr ago. This cannot be attributed to the different stellar evolutionary models, as the same BAsTI isochrones were used in our work. Moreover, none of the regions from \cite{delPino2013} show the prominent burst at $4.6 \pm 0.4$ Gyr ago, which could have been smoothed out by a shallower photometry. In comparison, the FWHM resolution in the present study is $\sim$0.2, 0.2, 0.3, 0.9, 2 Gyr at the lookback times $\sim$0.5, 0.9, 1.8, 4.6, 11.3 Gyr in the most crowded region Fornax1, while \cite{delPino2013} reported the values equivalent to FWHM of $\sim$1.4 Gyr at an age 3 Gyr ago and $\sim$4 Gyr at the oldest ages achieved by their photometry. Aside from the major differences, our results for Fornax1 and Fornax2 agree with the results from \cite{delPino2013} in terms of the declining $\psi(t)$ at increasing elliptical radii, as well as towards younger ages at all radii and in the total recovered metallicity range of $z(t)$.

\subsubsection{Metallicity distribution} \label{sec:comparison_b}

In independent spectroscopic studies of Fornax its metal enrichment was found to follow a radial gradient. The VLT/FLAMES data for 562 RGB stars from \cite{Battaglia2006} showed that the range of metallicity in a stellar sample located within $r<3'$ was $-1.4\lesssim$[Fe/H]$\lesssim-0.4$, while for $3'< r<13'$ metallicities within $-2.1\lesssim$[Fe/H]$\lesssim-0.2$ were measured. In comparison, the metallicity range in our solutions is $\rm -1.7<[Fe/H]<-0.4$ for both Fornax1 ($r\simeq3'$) and Fornax2 ($r\simeq13'$), and agrees with the previous results within the errors indicated in Figure~\ref{fig:sfhs}. Some of the most metal-poor stars with $\rm [Fe/H]=-1.7\pm0.1$ in the \cite{Battaglia2006} RGB sample had an inferred age $>10$ Gyr old. The metallicities of [Fe/H]$=-1.0\pm0.1$ and [Fe/H]$=-0.7\pm0.1$ were associated with 2-5 and 1.5-2 Gyr old isochrones. Our solutions from photometric data agree well with these estimates obtained with spectroscopic measurements, where each of the abundance levels mentioned above corresponds to the age of the onset of each of the major star formation episodes. Additionally, \cite{Battaglia2006} reported the presence of stars that were formed a few hundred Myr ago in the central regions (see also  \citealt{Mapelli2009} for evidence that these stars are genuine young stars and not blue straggler stars). This is also in agreement with the recovered SFH of the galaxy in our study. 

In Appendix~\ref{sec:mdf} we make a detailed comparison between the MDF of Fornax derived from the SFH, and that obtained from the spectroscopic studies by \citet{BattagliaStarkenburg2012} and \citet{Kirby2010}.  It is interesting that the MDF for Fornax1(2) derived in this paper is bimodal (with peaks at $\rm [Fe/H]\simeq -1.4$ and $\rm [Fe/H]\simeq -0.9$, see Figure~\ref{fig:mdf_noerror}), indicating that these regions have at least two different metallicity populations associated with the distinct star formation burts. This is in contrast with the spectroscopic MDFs which appear to be dominated mostly by the single peak around $\rm [Fe/H]\simeq -1.0$. In the Appendix, we demonstrate that simulating the spectroscopic metallicity errors in the metallicities estimated here from photometry smoothens out the bimodality, resulting in an MDF closely resembling the spectroscopic one (Figure~\ref{fig:mdf_error}, with possibly a difference in the metallicity zero-point.

Interestingly, populations with distinct spatial and chemo-dynamic properties that result in a decomposition of the spectroscopic MDF into several metallicity components, have been inferred in Fornax by previous studies. Based on the spectroscopy of \cite{Battaglia2006} with data covering the tidal radius, and using Bayesian analysis, \cite{Amorisco2012} separated the Fornax stellar populations into metal poor (MP; $\rm [Fe/H]\approx-1.60$), intermediate metallicity (IM; $\rm [Fe/H]\approx-0.95$) and metal rich (MR; $\rm [Fe/H]\approx-0.65$), which robustly fitted the properties of the stellar spatial distribution, metallicity and line of sight velocity. The MDFs of Fornax1(2) derived from our SFH (Figure~\ref{fig:mdf_noerror}) have two peaks centred around $\rm [Fe/H]\approx-1.40-(-1.35)$ and $\rm [Fe/H]\approx-0.90-(-0.74)$, which approximately agree with the MP and IM+MR populations inferred in the previous work. While it is hard to distinguish potential IM and MR components in our MDFs, we can only speculate that if the third population was present in our MDF solution, it would be likely more pronounced in Fornax1, at the centre of the galaxy, than in Fornax2, just outside of the core region. This agrees with the high-metallicity tails of the MDFs of Fornax1(2), as can be seen in the figure. Moreover, the three distinct episodes of star formation found in our solutions for Fornax1 (old, intermediate and young) support the hypothesis of three distinct spatial and chemo-dynamic populations of \cite{Amorisco2012}. \cite{Hendricks2014} calculated abundances of $\alpha$-elements and MDF of Fornax based on their spectroscopic CaT data at two outer regions in between one core and one tidal radii along the major axis, slightly extending the coverage of the \cite{Battaglia2006} sample with a similar selection of the RGB stars. They also identified three distinct sub-populations by fitting an analytic function. Although the sub-populations differed in metallicity from the ones in \cite{Amorisco2012}, their two most metal rich populations ($\rm [Fe/H] = -0.97$, $-1.45$) peaked at [Fe/H] similar to the two peaks in this work ($\rm [Fe/H]\approx-0.90-(-0.74)$ and $\rm [Fe/H]\approx-1.40-(-1.35)$), while their most metal poor population at $\rm [Fe/H] = -2.11$ end was almost absent from our solutions, given that the Fornax1(2) are dominated mostly by the intermediate-metallicity and metal rich stars.

Driven by the photometric data and results of CMD fitting from \cite{delPino2013}, the study by \cite{Piatti2014} investigated the MDF in their regions IC1, IC2 and OC (as on Figure~\ref{fig:footprints}) and observed bimodality in all, with the MDF peaks centred around $\rm[Fe/H] \approx -1.25$ and $\rm[Fe/H] \approx -0.95$ in the central field IC1, similarly to the MDFs presented here.

\section{Discussion. The bursty SFH: what did reignite star formation?} \label{sec:discussion}

Simulations of SFHs of isolated dwarf galaxies often predict episodic bursts of star formation owing to the effects of Supernova feedback \citep{Shen2014, Stinson2007}, which temporarily deplete the central regions of star-forming material. This process leads to quenching star formation as gas flows outward into a hot halo, where it cools until subsequently sinks to the centre of the potential well, triggering a new burst of star formation. Other models \citep[e.g.][]{BenitezLlambay2015, Verbeke2014} invoke the combined effects of cosmic reionization, supernova feedback and late gas accretion to explain major bursts of star formation in dwarf galaxies after some quiescent epoch following an initial star formation burst. The SFH of Fornax is not unlike some predictions by these simulations.

Among other possible scenarios, a merger of two galaxies is also expected to induce star formation. This has been studied in various simulations by \cite{Teyssier2010}, \cite{Cox2008}, \cite{diMatteo2008}, \cite{Mihos1994}, albeit for galaxies with mass much larger than Fornax. Indications of mergers were found in dwarf galaxies such as Andromeda II \citep{Amorisco2014NaturAndII} and Sextans \citep{Cicuendez2018}. Various features observed in Fornax also show that it may have experienced a merger event. \citet{Amorisco2012} review previous observations that hint at the presence of subpopulations in Fornax, indicated by age-dependent asymmetries in the isophotes of stellar populations \citep{Wang2019b, delPino2015, Stetson1998, Irwin1995, Eskridge1988, Hodge1961b}, stellar overdensities \citep{Bate2015, deBoer2013, Coleman2005, Coleman2004}, as well as multiple age and chemo-dynamic components \citep{Battaglia2006, Saviane2000}. Therefore, as it was hypothesised by \cite{Coleman2004} and further supported by the numerical simulations of the merger in \cite{Yozin2012}, based on proper motions available at the time, and considering the kinematic properties of different subpopulations in \cite{Amorisco2012}, the observed spatial distribution of the stellar populations could result from a merger of a bound pair of dwarf galaxies in a relatively recent period of time, $\sim2-3$ Gyr ago. Such scenario could be in agreement with the bursts starting to occur 2.8 and 1.9 Gyr ago in Fornax1 and continuing at a later time in Fornax2 after the shocks have reached the radius of this region\footnote{As noted before (Section~\ref{sec:sfh_results}), the bursts at $\lesssim$2 Gyr ago have a low statistical weight, therefore this is only a speculative statement assuming the bursts are true.}. Other evidence based on chemo-kinematic properties of Fornax has been used to propose a possible merger scenario $\sim6$ Gyr ago in \cite{delPino2017}. These predictions coincide with the ages of the onset of the intermediate-age star formation bursts in the SFH solutions for Fornax1 and Fornax2. Lastly, an early merger has also been suggested by \citet{Leung2020} to explain the unusually large mass of stars deposited in five GCs of Fornax, a mismatch between the metallicity of the GC5 stars and the old field stars and the differential dynamics of different stellar populations. In their scenario, Fornax would have merged with a $\sim$3 times less massive galaxy $\sim$10 Gyr ago. This hypothesis could explain what contributed to the oldest star formation episode at the median age $\sim10.6$ Gyr ago in our solutions for the central region Fornax1.

Additionally, there is theoretical and observational evidence that tidal interaction between a host and a dwarf galaxy may also ignite star formation in either of the two galaxies \citep[e.g.][for an early reference]{Larson1978}. Indications of tidally activated star formation after interaction with a smaller satellite galaxy have been found in a M31-M33 pair and our own MW interacting with the Sagittarius dSph. \citet{Bernard2012} showed that M31 experienced a burst of star formation in its disk outskirts that could have been triggered by a close approach of M33, reproduced by numerical simulations in \cite{McConnachie2009}. \cite{RuizLara2020a} have discussed recurrent episodes of enhanced star formation in the MW disk that coincide with independently inferred pericentric passages of the Sagittarius dSph galaxy \citep[e.g.][]{Purcell2011, delavega2015, Laporte2019}. In turn, the SFH of the Sagittarius core shows star formation enhancements at similar times \citep{Siegel2007}. Also, numerical simulations of the SFH and orbit of the Carina dSph \citep{Pasetto2011}, which has an extended SFH and well separated star formation episodes \citep{SmeckerHane1994, Monelli2003} similar to Fornax, showed correlation between the peaks in SFR and pericentric passages around the MW. Interestingly, the new orbits calculated by \cite{Patel_20} based on \textit{Gaia} proper motions and including the influence of the Magellanic Clouds indicate a second last pericentric passage around 5 Gyr ago when the influence of the Magellanic Clouds is taken into account, in agreement with the similar intermediate-age star formation burst of Carina observed in the SFH derived by \citet{Santana2016}. Finally, a close passage of Leo I about the MW (entering the virial radius of the MW $\simeq$ 2 Gyr ago, and pericentre at $91\pm36$ kpc about 1 Gyr ago, \citealt{Sohn_13}) may be linked with an enhancement of the Leo I star formation activity some 2 Gyr ago and lasting to about 1 Gyr ago \citep{RuizLara2020b}.

To study whether pericentric passages about the MW could be responsible for the Fornax bursty SFH, we calculated the orbit of Fornax in the potential of the MW.  For the properties of Fornax we use the \textit{Gaia} proper motion of \citet{Fritz_18} (but note that the proper motion of \citealt{Helmi_18} is very similar when compared with the systematic error of 0.035 mas/yr). Orbit integration is performed with the \textit{galpy}\footnote{Available at: \url{https://github.com/jobovy/galpy}.} package of \cite{Bovy_15}, where for the MW potential we use variants of \textit{MWPotential14}, with the halo mass modified to obtain MW masses which are more in-line with recent mass measurements. The values of m$_\mathrm{vir}=1.45\times 10^{12}$ M$_\odot$ and $1.1\times10^{12}$ M$_\odot$ are considered to explore the range of MW  masses \citep{Fritz_20,Bland_16}.

A preliminary investigation included potential encounters of Fornax with other galaxies brighter than ultra-faint dwarfs in the absolute luminosity-based definition of \citet{Simon_19a}. We found that within the past 1 Gyr, a close encounter of Fornax with one of the dwarf galaxies is unlikely. The closest passages within the past 2.5 Gyr are possible with Crater II (1.3 Gyr ago) and with Sculptor (1.6 Gyr ago); however, given the observational uncertainties, there is a chance of only $\sim10\%$ for the dwarfs to approach each other as close as 20 kpc, which significantly exceeds their stellar extent. These scenarios are not considered here in detail due to increasingly large errors at the lookback times longer than 1 Gyr. Therefore, we cannot constrain well any dwarf-dwarf encounters that could have caused the star formation bursts of Fornax.

\begin{figure}
	\includegraphics[width=\columnwidth]{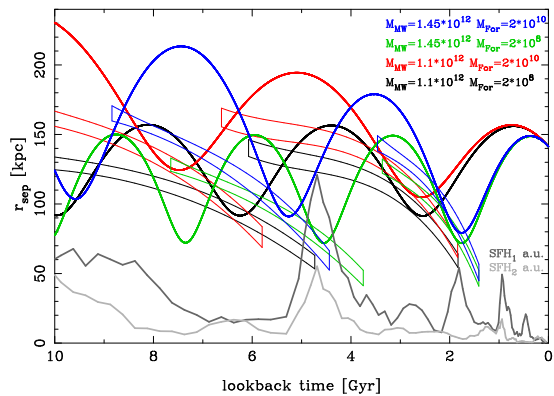}
	\caption{The figure shows the Galactocentric distance of Fornax over cosmic time. The thick lines use the most likely properties of Fornax for different masses of MW and Fornax. The thin lines outline the range of pericentre times and distances within the central 68.3\% interval of pericentre times due to proper motion and distance measurement errors. For comparison, the star formation history in both fields is shown in arbitrary units.}
    \label{fig:peri_passages}
\end{figure}

Our main analysis concerns the separation distance between Fornax and the MW. This orbit integration included dynamic friction and two mass estimates for Fornax that bracket most mass estimates in the literature. The lower value of $2\times10^{8} M_\odot$ represents the mass contained within the stellar extent of Fornax \citep{Diokogiannis_17} and the higher mass of $2\times10^{10} M_\odot$ was obtained by \citet{Read_19} by using a cored, untruncated NFW halo. The results are similar to the ones in \citet{Fritz_18} and are shown in terms of separation distance $\rm r_{sep}(t)$ as a function of lookback time in Figure~\ref{fig:peri_passages}.

In the models with the more massive MW, the two most recent inferred pericentric passages correspond to the two most recent substantial bursting epochs in Fornax1 that occurred $1.8\pm0.2$ and $4.6\pm0.4$ Gyr ago. For the higher mass of both galaxies, which are considered most likely, the last pericentre occurred $1.7^{+1.7}_{-0.3}$ Gyr ago at a distance of $76^{+66}_{-24}$ kpc and the second latest was $5.1^{+3.6}_{-0.7}$ Gyr ago at $87^{+75}_{-28}$ kpc. The timing of the passage $\sim5.1$ Gyr ago coincided with the onset of the most intense burst, which ceased in the next $\sim0.5-1$ Gyr. The difference of ages of the two passages is more similar to the age difference of the two considered bursts when the low Fornax mass is adopted. In this case, both bursts occur exactly at the pericentre, while for the large Fornax mass, either passage is off by 0.3-1 Gyr.

Note that, given the measurement errors, a wide range of times for the pericenter passages is possible (as indicated by the boxes depicted with thin lines in Figure~\ref{fig:peri_passages}). Additionally, the calculations included the assumption of a spherical halo \citep{Bovy_16}, while improved simulations should account for deviations from spherical symmetry (the most prominent effects could be due to the LMC). These considerations would produce somewhat different results, as shown in \cite{Erkal_19a}, \cite{Sohn_17}. In fact, the former study found a probability of Fornax once being a satellite of the LMC, and possible effects of such interaction should also be considered. Very recently, \citet{Patel_20} performed orbit calculations including both Magellanic Clouds and it was found that although Fornax is not bound to the Clouds, their effect changed passage times of Fornax, as compared to our results. They show the most recent peri-centric passages at $\sim$1.5 and 5 Gyr ago for their more massive MW model (see Figure~2 therein).

The latter and our calculations obtain pericentric passages that coincide with significant bursts of star formation in the second half of Fornax history. It must be noted that more stringent constraints on the mass of the MW and shape of its potential can change the inferred dating of the pericentric passages drastically. Therefore, the only unambiguous conclusion from this evidence follows that if the tidal interactions are assumed to be the mechanism that triggered recurrent star formation in Fornax, then this would be likely if the galaxies had the masses $M_{MW}=1.45\times10^{12}M_{\odot}$ and $M_{For}=2\times10^{10}M_{\odot}$ out of the four considered mass combinations. If true, this would provide tantalising evidence that the MW tidal effects on Fornax at a distance of $\simeq$ 80 kpc may be the cause of these bursts. This distance, in fact, places Fornax well within the MW virial radius at the time of the tidal interaction.

If the assumption is confirmed, then \cite{Battaglia2015} conclusion that tidal effects in Fornax were insignificant and did not disturb its stellar or dark matter distributions provides support to the possibility that neither ram pressure nor tidal stripping have been able to remove the gas that may still have been gravitationally bound to Fornax after the first epoch of star formation, but rather, may have tidally induced star formation.

\section{Summary} \label{sec:summary}

We derived the SFH in two small fields near the centre of the Fornax dSph with much deeper photometric data than used previously, which resulted in a significantly improved age resolution, sufficient to date individual episodes of star formation and estimate their intrinsic duration. This work is thus providing an unprecedented sharp view of the evolutionary history of the Fornax stellar populations.

The work is concentrated on two HST/ACS regions of Fornax ($\sim3.4\times3.4$ squared arcmin each) at radial distances of $2.8'$ and $12.7'$ from its centre. We employed the CMD fitting technique to analyse the stellar population composition of the galaxy and derived its SFH. Careful tests using mock stellar populations have allowed us to put tight limits on the duration of the bursts, and on the possibility of low-level star formation between the bursts. We have found that the first and most abundant populations of old, metal-poor stars, accounting for up to $\sim70\%$ of stars in Fornax1 and $\sim80\%$ in Fornax2, were formed in a star formation episode $\sim10-11$ Gyr ago, that lasted between $\sim4-6$ Gyr in Fornax1 and up to $\sim4.25$ Gyr in Fornax2. The second strong and sharp burst $\sim4.6$ Gyr ago contributed $\sim10-15\%$ to the mass in stars during 1 Gyr at most in both fields. Finally, the subsequent significant episode $\sim1.8$ Gyr ago in Fornax1 and younger intermittent bursts in both fields added the remaining mass in stars, with only $1-3\%$ of it contained in the stars formed less than 1 Gyr ago. Due to the small number of young stars, we cannot claim accurate ages or metallicities for the recent star formation activity. The results of the mock stellar population tests indicate that both regions (and more securely Fornax1) have likely ceased star formation activity between $\sim8-5.5$ Gyr ago, while low level star formation seems to have occurred between the intermediate-age and the young bursts.

The SFHs for two regions of such small size are probably not representative of the whole core region or the whole galaxy, given the non-uniform distribution of different stellar populations. However, the main star formation bursts are observed in both regions, with the same age, similar inferred duration and relative burst intensity. This is a further indication of the reliability of the derived scenario. 

The two regions also seem to be connected by the spatial age gradient, typical for dwarf galaxies, where the central region Fornax1 tends to experience more prolonged star formation episodes than Fornax2. It is very remarkable that 
this  outside-in quenching of star formation, commonly observed in dwarf galaxies \citep[e.g.][and references therein]{Hidalgo2013}, is observed in both main epochs of star formation: both the old burst and the young star formation activity have a longer duration in the innermost field. That is, if star formation would not have been reignited in Fornax at intermediate ages, the galaxy would still show the usual stellar population gradient observed in dwarf galaxies.

We have discussed the possible mechanisms that may be responsible for the separated events of star formation in Fornax. Numerical simulations often predict a bursty SFH in field dwarf galaxies, and thus the same characteristic in Fornax could be just an example of such behaviour owing to the normal secular evolution of this dwarf galaxy. A second possibility is that it is related to a dwarf-dwarf merger: indications of an early one have been discussed in \citet{Leung2020} and would explain the peculiar Fornax GC population, while a more recent merger could explain the multiple chemo-dynamic and age components \citep{Amorisco2012}. We discuss a third possibility, in relation to possible repeated tidal interactions of Fornax with the MW. We show that, for the mass of the MW favoured by the most recent determinations, and masses of Fornax within current (wide range) estimates, the most likely times for the two most recent pericenter passages of Fornax are in very good agreement with the times of the intermediate-age and main young star formation burst in Fornax. This adds another example of tidally induced star formation to the mounting evidence of such occurrence, which can be particulary well characterized in the LG, where not only current, but also intermediate-age and old star formation events can be accurately dated. 

As a final remark, we would like to note that this work demonstrates that ultra-precise CMDs of LG galaxies, such as the ones that will be obtained by the LSST project for dwarfs in the MW system, will allow a much sharper view of the details of their SFHs across all epochs of their evolution, including accurate times of the star formation events  and their duration. Such studies of LG galaxies, together with complementary information coming from other present and upcoming galactic archaeology surveys such as Gaia, will significantly boost our knowledge of evolution profiles of galaxies leading to the LG as we currently observe it.

\section*{Acknowledgements}

We would like to thank the anonymous Referee for the detailed report and insightful comments, which we believe greatly helped to improve the quality of the manuscript. We thank Giuseppina Battaglia for kindly providing their spectroscopic catalogues and useful discussions, and Andr\'es del Pino for providing the results of their SFH work on Fornax. This work has made use of data stored at Barbara A. Mikulsi Archive for Space Telescopes (STScI), NASA Astrophysics Data System Bibliographic Services and the NASA/IPAC Extragalactic Database (NED), as well as ESO Science Archive Facility. Used facilities: HST (ACS).

TKF, CG, MM and TRL acknowledge support through Ministerio de Econom\i a, Industria y Competitividad  and AEI/FEDER (UE) grant AYA2017-89076-P, and by the Consejer\'\i a de Econom\'\i a, Industria, Comercio y Conocimiento of the Canary Islands Autonomous Community, through the Regional Budget (including IAC projects "Galaxy Evolution in the Local Group" and "TRACES"). TRL additionally acknowledges financial support through grants AYA2016-77237-C3-1-P (RAVET project), a Juan de la Cierva - Formaci\'on grant (FJCI-2016-30342) and support from a Spinoza grant (NWO) awarded to A. Helmi, grants (AEI/FEDER, UE) AYA2017-89076-P, AYA2015-63810-P, AYA2016-77237-C3-1-P (RAVET project), and a MCIU Juan de la Cierva - Formaci\'on grant (FJCI-2016-30342). SC acknowledges support from Premiale INAF MITiC, from INFN (Iniziativa specifica TAsP),  PLATO ASI-INAF contract n.2015-019-R0 \& n.2015-019-R.1-2018.

\textit{Software:} \verb|DAOPHOT| \citep{Stetson1987,Stetson1994}, synthetic CMD code (thanks to Santi Cassisi), \verb|TheStorm| \citep{Bernard2018}, \verb|numpy| \citep{numpy2020}, \verb|scipy| \citep{scipy2020}, \verb|astropy| \citep{astropy2018,astropy2013}, \verb|matplotlib| \citep{Hunter2007}, \verb|pandas| \citep{mckinney2010}.

\section*{Data Availability}

The data used in this article are available in Barbara A. Mikulsi Archive for Space Telescopes at \url{http://archive.stsci.edu/}. The data is part of the proposal 13435 (PI: M. Monelli). The text of the proposal is available at: \url{http://archive.stsci.edu/proposal_search.php?id=13435&mission=hst}.





\bibliographystyle{mnras}
\bibliography{references} 




\appendix

\section{HST/ACS Observations Log} \label{sec:hst_log}
Table \ref{tab:datalog} details the HST/ACS observations that were used in this paper.

\begin{table*}
 \caption{Log table indicating the images taken as part of HST/ACS proposal ID 13435 used in this paper. Columns (left to right): names of image files, target of observations indicated in the image header, date and time of observations, right ascension and declination coordinates (J2000), ACS filter, exposure time, modified Julian date. Top half of data is for Fornax1, bottom - for Fornax2.}
 \label{tab:datalog}
 \begin{tabular}{lcccccccc}
  \hline
  Image   &   Target     & Date       & Time     & RA($^h:^m:^s$)    & DEC($\degr:':''$)    & Filter & Exposure (s) & Julian \\
  \hline
  jcb009c7q\_flc.fits & Fornax1 & 2014-01-04 & 03:07:04 & 02:39:42.55 & -34:31:43.25 & F475W  & 80  & 5.66611E+04 \\
  jcb009c9q\_flc.fits & Fornax1 & 2014-01-04 & 03:54:37 &   02:39:42.43 & -34:31:44.58 & F814W  & 850 & 5.66611E+04 \\
  jcb009cfq\_flc.fits & Fornax1 & 2014-01-04 & 04:21:06 &   02:39:42.55 & -34:31:43.25 & F475W  & 520 & 5.66611E+04 \\
  jcb010clq\_flc.fits & Fornax1 & 2014-01-04 & 05:37:33 &   02:39:42.55 & -34:31:43.25 & F814W  & 80  & 5.66612E+04 \\
  jcb010cnq\_flc.fits & Fornax1 & 2014-01-04 & 05:42:47 &   02:39:42.43 & -34:31:44.58 & F475W  & 850 & 5.66612E+04 \\
  jcb010ctq\_flc.fits & Fornax1 & 2014-01-04 & 06:09:14 &   02:39:42.55 & -34:31:43.25 & F814W  & 520 & 5.66612E+04 \\
  jcb011v7q\_flc.fits & Fornax1 & 2013-12-27 & 14:21:06 &   02:39:42.55 & -34:31:43.25 & F814W  & 80  & 5.66535E+04 \\
  jcb011v9q\_flc.fits & Fornax1 & 2013-12-27 & 14:26:20 &   02:39:42.43 & -34:31:44.58 & F475W  & 850 & 5.66536E+04 \\
  jcb011vfq\_flc.fits & Fornax1 & 2013-12-27 & 14:52:47 &   02:39:42.55 & -34:31:43.25 & F814W  & 520 & 5.66536E+04 \\
  \hline
  jcb006d2q\_flc.fits & Fornax2            & 2014-06-22 & 03:34:03 &   02:40:39.53 & -34:32:41.60 & F814W  & 80  & 5.68301E+04 \\
  jcb006d4q\_flc.fits & Fornax2            & 2014-06-22 & 03:39:17 &   02:40:39.65 & -34:32:40.27 & F475W  & 850 & 5.68301E+04 \\
  jcb006daq\_flc.fits & Fornax2            & 2014-06-22 & 05:05:41 &   02:40:39.53 & -34:32:41.60 & F814W  & 520 & 5.68302E+04 \\
  jcb007fhq\_flc.fits & Fornax2 & 2014-06-22 & 22:26:02 &   02:40:39.53 & -34:32:41.60 & F475W  & 80  & 5.68309E+04 \\
  jcb007fjq\_flc.fits & Fornax2 & 2014-06-22 & 22:31:15 &   02:40:39.65 & -34:32:40.27 & F814W  & 850 & 5.68309E+04 \\
  jcb007fpq\_flc.fits & Fornax2 & 2014-06-22 & 22:57:44 &   02:40:39.53 & -34:32:41.60 & F475W  & 520 & 5.68309E+04 \\
  jcb008b9q\_flc.fits & Fornax2 & 2014-06-21 & 20:57:11 &   02:40:39.53 & -34:32:41.60 & F814W  & 80  & 5.68298E+04 \\
  jcb008bbq\_flc.fits & Fornax2 & 2014-06-21 & 21:02:25 &   02:40:39.65 & -34:32:40.27 & F475W  & 850 & 5.68298E+04 \\
  jcb008bhq\_flc.fits & Fornax2 & 2014-06-21 & 21:28:52 &   02:40:39.53 & -34:32:41.60 & F814W  & 520 & 5.68298E+04 \\
  \hline

 \end{tabular}
\end{table*}

\section{Metallicity Distribution Function} \label{sec:mdf}

This Appendix describes the comparison of the MDF of Fornax obtained from the SFH retrieval procedure and from some of the previous spectroscopic studies of Fornax.

Among the considered spectroscopic samples with the spatial coverage including Fornax1(2) (medium resolution \citealp{Kirby2010}; medium and high resolution \citealp{Letarte2010}; Ca II triplet spectroscopy from \citealp{BattagliaStarkenburg2012}), the samples from \cite[][hereafter, K10, B12]{Kirby2010,BattagliaStarkenburg2012} were used as the most numerous. Since the number of stars measured spectroscopically in the Fornax1 and Fornax2 fields is small, for the comparison we are using larger samples of stars within the Fornax central region. Although no substantial stellar population gradients are expected in the Fornax core that can affect the metallicity distribution, particularly of RGB stars, which are older than $\simeq 2$ Gyr \citep{delPino2015}, the spatial variation of the MDF samples is investigated here.

\begin{figure}
    \includegraphics[width=\columnwidth]{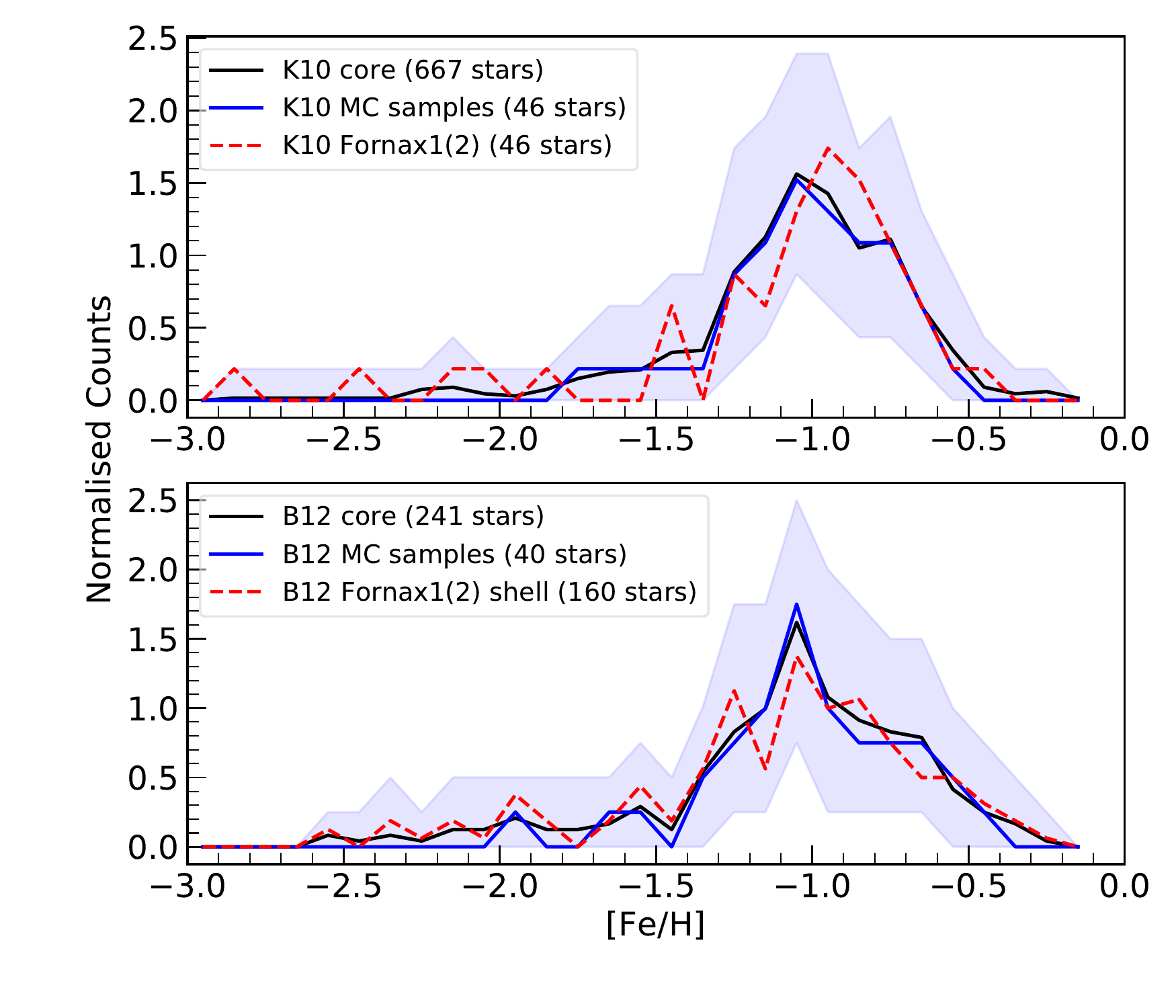}
    \caption{Normalised MDFs of Fornax1+2 and Fornax core for the K10 (top panel) and B12 (bottom panel) spectroscopic datasets. Black solid lines represent the MDF of the Fornax core; blue solid line is the median MDF of 1000 Monte Carlo draws of 46 stars from the core sample with the 90\% confidence interval indicated by the blue error margin; red dashed line shows the MDF of stars within Fornax1+2 only.}
    \label{fig:mc_samples}
\end{figure}

To test how the MDFs can be affected by random sampling of stars within an elliptical radius equal to that of Fornax2, 1000 Monte Carlo samples (46 stars each, which is similar to the number of stars spectroscopically measured in Fornax1+2 by K10) were drawn from the full spectroscopic samples. Figure~\ref{fig:mc_samples} shows, for K10 (top) and B12 (bottom), the MDF of the stars measured within the core radius (black line), the median MDF from the Monte Carlos samples (blue line) and the MDF of stars measured in the fields Fornax1+2 (for K10) or the stars measured within elliptical shells covering the regions Fornax1 and Fornax2 (for B12, as only 6 stars from their sample coincided with the exact regions Fornax1 and Fornax2). The median of the Monte Carlo samples is consistent with the MDF samples of Fornax1+2 within the 90\% confidence interval, meaning that the metallicity data are not expected to have a significant spatial bias.

\begin{figure}
    \includegraphics[width=\columnwidth]{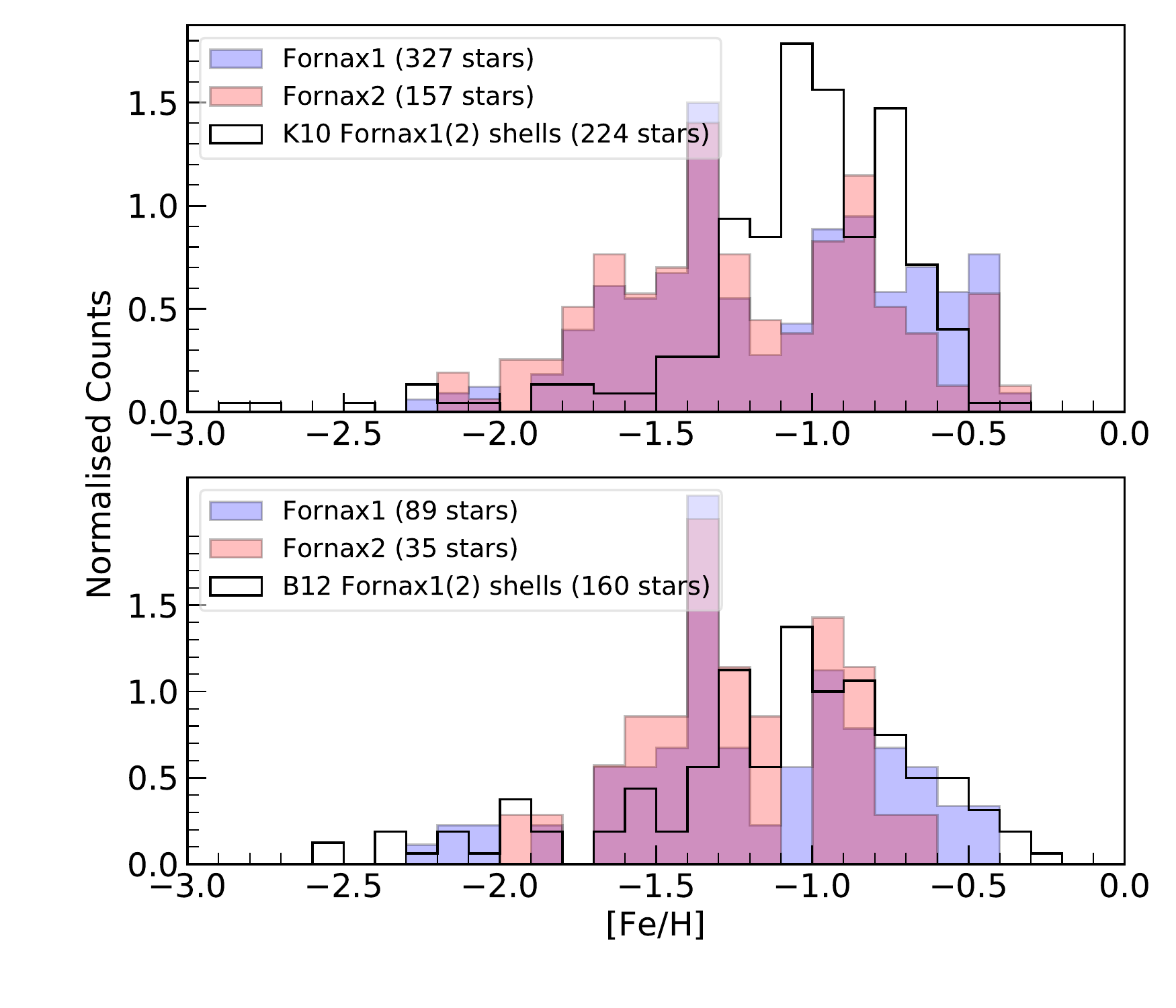}
    \caption{Normalised MDFs of Fornax1+2 and Fornax core. Top panel shows the K10 dataset: blue (Fornax1) and red (Fornax2) histograms show MDFs obtained from the SFH of the regions and black histogram shows the MDF from K10 for both Fornax1 and Fornax2. Bottom panel is the same as the top but for B12.}
    \label{fig:mdf_noerror}
\end{figure}

In spite of the fact that the results of the conducted test showed little dependence of the MDF on spatial positions of stars, the spectroscopic data used for the analysis below were selected from the elliptical shells enclosing Fornax1 and Fornax2 and not from the whole core to further reduce the possibility of a spatial bias. Figure~\ref{fig:mdf_noerror} illustrates the comparison of the MDF derived from the SFH solutions in this work and the MDF from spectroscopic samples with similar number of stars. The photometric selection criteria for the stars in the solution CMD were adjusted according to the spectroscopic data sets: K10 includes stars just above the RC ($M_{F814W}<-0.7$ according to Figure~\ref{fig:obs_cmds}); B12 covers the $\sim$2 brightest magnitudes of the RGB only. As can be seen from Figure~\ref{fig:mdf_noerror}, Fornax1 and Fornax2 MDFs derived from the SFH are similar along most of the metallicity range, with Fornax1 being slightly more metal-rich. Both are characterised by a bimodal MDF, with a metal-poor peak at $\rm [Fe/H]\approx-1.4$ and a metal-rich peak at $\rm [Fe/H]\approx-0.9$, which correspond with the stellar populations that are $\sim6-14$ Gyr old (Fornax1) and $8-14$ Gyr old (Fornax2) and $\sim 0-6$ Gyr old (Fornax1, 2), respectively. This distribution is unlike the K10 and B12 counterparts, which peak at $\rm [Fe/H]\approx-1.0$ and are skewed towards the more metal-rich end. Another clear disagreement, particularly with K10, occurs at $\rm -1.8 < [Fe/H] < -1.3$, with this range represented by a larger number of stars in Fornax1(2).

The MDF showing at least two distinct modes in this work is consistent with the derived SFH, characterised by major old and intermediate bursts of star formation, likely separated by an epoch of lower activity around 6-8 Gyr ago, and metallicity increasing with time as a consequence of star formation. At this point, we consider the following question: if the true MDF of Fornax was indeed bimodal, could the measurement errors of the spectroscopic metallicities erase the bimodality?

\begin{figure}
    \includegraphics[width=\columnwidth]{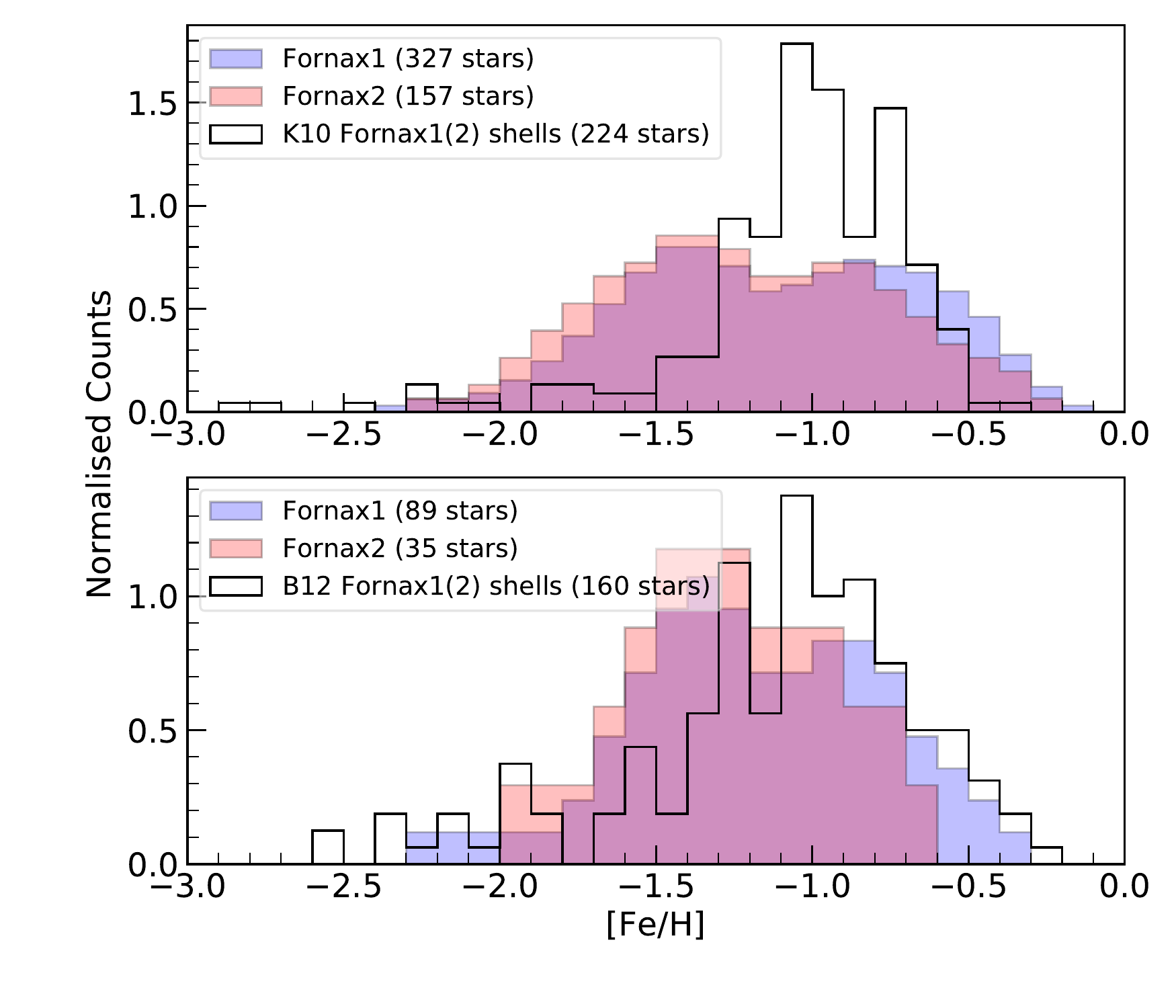}
    \caption{The same as Figure~\ref{fig:mdf_noerror} but with the Fornax1(2) MDFs smoothed with the uncertainties of K10 and B12 metallicities.}
    \label{fig:mdf_error}
\end{figure}

To answer this question, we simulated the spectroscopic errors (as drawn from the K10 and B12 data tables\footnote{The spectroscopic metallicities presented in B12 were kindly made available by G. Battaglia.}) in the MDFs derived from the SFH. The results are shown in Figure~\ref{fig:mdf_error}, where the Fornax1(2) MDFs represent medians of 1000 randomly drawn Gaussian error samples ($\sigma \approx 0.14$ dex in K10 and B12) combined with the MDFs from the SFH. This step resulted in smoothing of the Fornax1(2) distributions and partial resolution of the bi-modality disagreement, particularly in comparison with B12.

With regards to the second disagreement with the spectroscopy, the metal-poor stars at $\rm -1.8 < [Fe/H] < -1.3$ are represented more in the Fornax1(2) solutions even after introducing the errors, as can be seen in Figure~\ref{fig:mdf_error}. The difference is more prominent in the case of K10.
Note that K10 warn about a possible bias in their sample because some stars on the extreme blue end of the RGB (populated mainly by asymptotic giant branch, extremely young or extremely metal-poor red giant stars) did not pass the spectroscopic selection criteria. This bias could underestimate the fraction of metal poor stars. Finally, it is possible that a zero-point in the metallicity scale could account for this remaining difference.

From the above analysis we conclude that the MDF of the solutions presented in this work is reasonably consistent within the errors with the MDF spectroscopically derived for the core region of Fornax, with a possible zero-point difference. The analysis shows that the MDF derived for Fornax1(2) from the fit to the CMD may have a higher resolution in metallicity than achieved by the K10 and B12 data sets, revealing a possible bimodality in the metallicity distribution of the stars in the core of Fornax.


\bsp	
\label{lastpage}
\end{document}